\documentclass{emulateapj}

\shorttitle{Rapid Disk-Jet Interaction in GRS~1915+105}
\shortauthors{Rothstein, Eikenberry \& Matthews}

\begin{document}

\submitted{Accepted for publication in The Astrophysical Journal}

\title{Observations of Rapid Disk-Jet Interaction in the Microquasar
GRS~1915+105}

\author{David M. Rothstein,\altaffilmark{1} Stephen
S. Eikenberry,\altaffilmark{2} and Keith Matthews\altaffilmark{3}}

\altaffiltext{1}{Department of Astronomy, Cornell University, 514
Space Sciences Building, Ithaca, NY 14853, droth@astro.cornell.edu}
\altaffiltext{2}{Department of Astronomy, University of Florida, 211
Bryant Space Science Center, Gainesville, FL 32611,
eiken@astro.ufl.edu} \altaffiltext{3}{California Institute of
Technology, Downs Laboratory, MS 320-47, Pasadena, CA 91125,
kym@caltech.edu}

\begin{abstract}

We present evidence that \mbox{$\sim 30$} minute episodes of jet
formation in the Galactic microquasar GRS~1915+105 may sometimes
entirely be a superposition of smaller, faster phenomena.  We base
this conclusion on simultaneous X-ray and infrared observations in
July 2002, using the {\it Rossi X-ray Timing Explorer} and the Palomar
5 meter telescope.  On two nights, we observed quasi-periodic infrared
flares from GRS~1915+105, each accompanied by a set of fast
oscillations in the X-ray light curve (indicating an interaction
between the jet and accretion disk).  In contrast to similar
observations in 1997, we find that the duration of each X-ray cycle
matches the duration of its accompanying infrared flare, and we
observed one instance in which an isolated X-ray oscillation occurred
at the same time as a faint infrared ``subflare'' (of duration
\mbox{$\sim 150$} seconds) superimposed on one of the main flares.
From these data, we are able to conclude that {\it each} X-ray
oscillation had an associated faint infrared flare and that these
flares blend together to form, and entirely comprise, the \mbox{$\sim
30$} minute events we observed.  Part of the infrared emission in 1997
also appears to be due to superimposed small flares, but it was
overshadowed by infrared-bright ejections associated with the
appearance of a sharp ``trigger'' spike in each X-ray cycle that were
not present in 2002.  We also study the evolution of the X-ray
spectrum and find significant differences in the high energy power law
component, which was strongly variable in 1997 but not in 2002.  Taken
together, these observations reveal the diversity of ways in which the
accretion disk and jet in black hole systems are capable of
interacting and solidify the importance of the trigger spike for large
ejections to occur on \mbox{$\sim 30$} minute timescales in
GRS~1915+105.

\end{abstract}


\section{Introduction}

GRS~1915+105 is one of the most fascinating objects in astrophysics
today.  It was one of the first systems in the Galaxy to be identified
as a ``microquasar'' --- an X-ray binary with relativistic jets which
mimics some of the behavior of quasars on a smaller and closer scale.
\citetext{For a review of these objects, see \citealp{Mirabel99} and
\citealp{Fender2004}; for a review focusing on GRS~1915+105, see
\citealp{FenderBelloni}.}  Because timescales in microquasars are
expected to be a factor of \mbox{$\sim 10^8$} shorter than in quasars
(owing to the much smaller mass of the compact object which powers the
system), microquasars are excellent laboratories for investigating
accretion disk evolution and jet formation in black hole systems.

GRS~1915+105 was first discovered as a transient X-ray source
\citep{Castro}, and its radio and infrared counterparts have since
been identified \citep{Mirabeletal94}.  The system contains a
\mbox{$\sim 14 \: M_{\sun}$} black hole fed by Roche lobe overflow
from a K-M giant companion \citep{GreinerMass,GreinerComp}.
GRS~1915+105 displays extreme variability in many different wavebands
and on many different timescales, making it unique among the X-ray
binaries.  Its uniqueness is likely due to its extremely high
accretion rate, which allows it to regularly reach luminosities unseen
in other Galactic X-ray sources \citep{Done2004,FenderBelloni}.

Perhaps the most spectacular variability observed from GRS~1915+105 is
seen in high resolution radio maps, which occasionally reveal the
presence of resolved, bipolar ejections with flux densities up to
\mbox{$\sim 600$} mJy that form on timescales of weeks and move away
from the system at relativistic speeds \citep[\mbox{$> 0.9
c$};][]{Mirabel94,RodMirabel99,Fender99}.  These events are referred
to as ``class A'' ejections by \citet{Eiken2000} to distinguish them
from smaller ``class B'' radio and infrared flares on \mbox{$\sim 30$}
minute timescales which are also thought to correspond to jet
ejection.  In addition to the radio and infrared flares, GRS~1915+105
displays a broad range of variability in the X-rays, where it switches
between many different states \citep[reviewed extensively
by][]{Belloni2000} and where quasi-periodic oscillations (QPOs) on
timescales as fast as \mbox{$\sim 168$} Hz have been observed
\citep{RemillardCargese,McClintock}.

Multiwavelength observations of ``class B'' events in GRS~1915+105 by
\citet{PooleyFender97}, \citet{Eiken98,Eiken98b}, and
\citet{Mirabel98} were the first to reveal the intimate link between
accretion disk evolution and relativistic jet formation on short
timescales in any black hole system.  \citet{Eiken98} observed a
one-to-one correspondence between repeating X-ray variability cycles
and infrared flares (\mbox{$\sim 100$} mJy dereddened) on timescales
of \mbox{$\sim 30$} minutes.  The observations are consistent with a
picture in which emptying and refilling of the X-ray emitting inner
disk \citep[e.g.][]{Belloni97a,Belloni97b} coincides with the ejection
of material into a jet, which radiates through synchrotron emission to
produce a flare \citep{Fender97,PooleyFender97}.  The infrared and
X-ray light curves decouple as the ejecta becomes causally separated
from the inner disk \citep{Eiken98}, and as the plasma moves out of
the accretion disk plane it can radiatively pump emission lines
originating within the disk \citep{Eiken98b}.  Each infrared flare
appears to be accompanied by a delayed radio flare, perhaps indicating
adiabatic expansion of the ejected cloud \citep{Mirabel98} or motion
along a conical jet \citep{FenderPooley98}, while the jet itself is
resolved by the Very Long Baseline Array as an optically thick
synchrotron source of length \mbox{$\sim 20$} AU that is variable on
similar timescales as the X-rays and infrared \citep{Dhawan}.

Though subsequent observations have revealed smaller, more complex
infrared flaring behavior in this source \citep[e.g.][]{Eiken2000},
the above represents the basic picture for the ejection of ``class B''
jets in GRS~1915+105.  In this paper, we present multiwavelength
observations of GRS~1915+105 on four nights during July 2002, when the
source was undergoing another period of ``class B'' ejection.  The
similarities and differences between these observations and those
obtained previously allow us to begin to map out how the many
different X-ray states of GRS~1915+105 affect jet production in this
unusual source.  We conclude that the observed ejections may sometimes
entirely be a superposition of smaller, more complex phenomena, and
that the appearance of a ``trigger'' spike in the X-ray light curve
\citep[seen in the observations of][but not in our July 2002
observations]{Eiken98} is a key ingredient for large, infrared-bright
ejections to occur on \mbox{$\sim 30$} minute timescales.

\section{Observations}

We obtained infrared observations of GRS~1915+105 on the nights of
2002 July 27--29 UT in the K ($2.2 \micron$) band, using the D-78
camera at the Cassegrain focus of the Palomar Observatory 5 meter Hale
telescope.  X-ray observations (with coverage between \mbox{$\sim$
2--100} keV) were obtained on July 27--28 and July 30 using the
Proportional Counter Array (PCA) on the {\it Rossi X-Ray Timing
Explorer} (RXTE); further details regarding this instrument can be
found in \citet{Greiner96} and references therein.

\begin{figure}
\epsscale{1.15}
\plotone{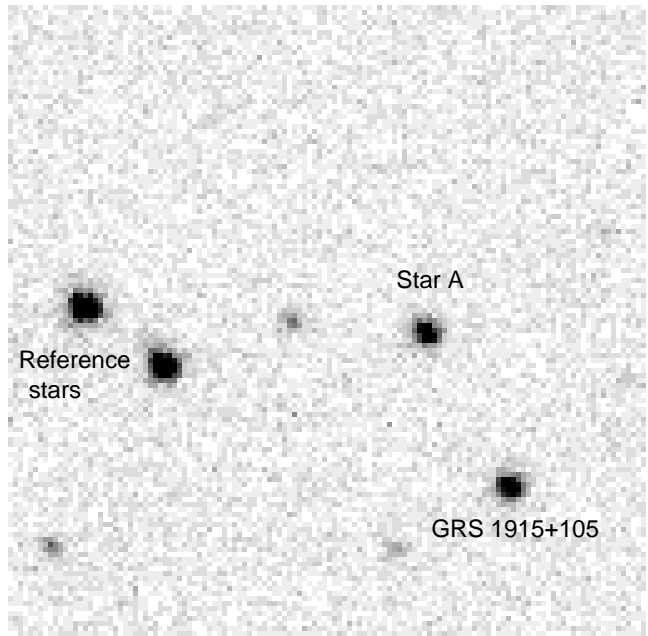}
\epsscale{1}
\caption{A typical $2.2 \micron$ infrared image of the field of
GRS~1915+105 at 1-second time resolution, from our observations on
2002 July 27.  North is up, east is to the left, and the field of view
is \mbox{$16 \arcsec \times 16 \arcsec$}.  The two reference stars
were used to correct the GRS~1915+105 and Star A light curves for
atmospheric variability and changes in the point spread function.}
\label{fig:grsim}
\end{figure}

We configured the infrared camera to take \mbox{$128 \times 128$}
pixel (\mbox{$16\arcsec \times 16\arcsec$}) images at a rate of one
frame per second, with absolute timing provided by a WWVB radio signal
from the National Institute of Standards and Technology (\mbox{$\sim
1$} ms accuracy).  We observed GRS~1915+105 in this mode for
\mbox{$\sim 8$} hours on each of the first two nights and for
\mbox{$\sim 1$} hour on July 29.  We processed each image by
subtracting an averaged sky frame, dividing by a flat field,
interpolating over bad pixels and filtering it in the Fourier domain
to remove electronic pattern noise that corrupted many of the images.
On July 27, we also observed the faint HST/NICMOS standard GSPC~P182-E
\citep[star no. 9177 from][]{Persson} for absolute flux calibration.

\begin{figure*}
\epsscale{1.14} \plotone{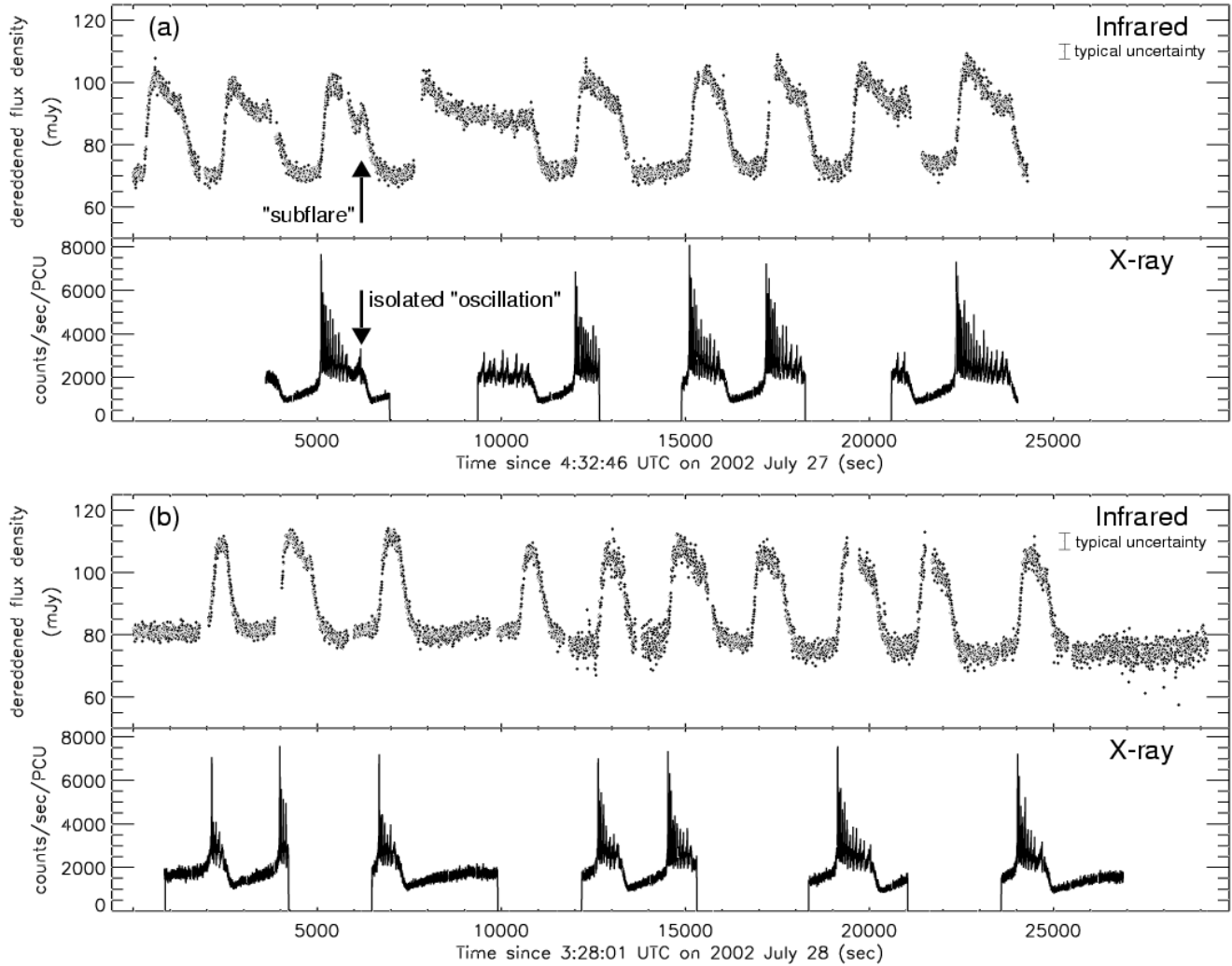} \epsscale{1}
\caption{Simultaneous infrared and X-ray light curves of GRS~1915+105
on (a) 2002 July 27 and (b) 2002 July 28.  Both light curves are at
1-second resolution, but the infrared data have been smoothed to
5-second resolution in this plot.  Typical 1-second \mbox{$\pm$ $1
\sigma$} uncertainties are shown for the infrared; the X-ray
uncertainties are assumed to be Poissonian and are too small to be
seen on this scale.  The infrared data were taken in the K ($2.2
\micron$) band and have been dereddened by 3.3 magnitudes, while the
X-ray data are in the \mbox{$\sim$ 2--100} keV energy range and are
plotted in units of counts per second per Proportional Counter Unit
(PCU) of the RXTE PCA instrument.  Gaps in the X-ray light curves
represent times when RXTE was not observing the source.  The
``subflare'' and isolated ``oscillation'' in (a) are discussed in \S
\ref{correlation}.}
\label{fig:lcurve}
\end{figure*}

A typical processed image is shown in Figure \ref{fig:grsim}.  The
field of view is large enough to capture GRS~1915+105 and several
nearby stars.  We performed differential photometry on GRS~1915+105
and the surrounding field stars, measuring their fluxes within a
\mbox{$1 \arcsec$} radius software aperture.  We used the two
brightest field stars as a reference, averaging their light curves
together and dividing the normalized result into the GRS~1915+105
light curve to correct for atmospheric variability and changes in the
point spread function.  This procedure was also applied to ``Star A''
in Figure \ref{fig:grsim}; its flux density remained steady at a value
of 3.1 mJy \citep[consistent with the reported value in][]{Fender97}
and was used to determine the flux density of GRS~1915+105.  The
GRS~1915+105 light curve was then dereddened by $A_{K}=3.3$ magnitudes
to correct for absorption in the Galactic plane \citep{Fender97}.  We
chose $A_{K}=3.3$ to be consistent with the previous literature,
although more recent work suggests that $A_{K}=2.2$ is a better
estimate for this source \citep{fuchs,Chapuis}.  If $A_{K}=2.2$ is
adopted, then all the infrared flux densities in this paper should be
reduced by a factor of \mbox{$\sim 3$}.

For the X-ray analysis, we extracted light curves (at 1-second time
resolution) from PCA {\tt Standard-1} data using FTOOLS v5.2.   We
also extracted X-ray spectra in the \mbox{$\sim$ 3--25} keV range,
using data from PCA Binned mode {\tt B\_8ms\_16A\_0\_35\_H\_4P} and
Event mode {\tt E\_16us\_16B\_36\_1s}.  We used standard procedures
for response matrix generation, background estimation and subtraction,
and correction for PCA deadtime.  We then used XSPEC v11.2 to fit each
spectrum with a standard model for black hole candidates consisting of
a ``soft'' component (which peaks in the low energy X-rays) modeled as
a multitemperature disk blackbody \citep[e.g.][]{Mitsuda} and a
``hard'' component (which extends to the higher energy X-rays) modeled
as a power law, both modified by hydrogen absorption fixed to a column
density of \mbox{$6 \times 10^{22}$} atoms cm$^{-2}$ \citep{Muno1999}.
A systematic error of 1\% was added to each spectrum before the fit
was performed.

\section{Light Curves}

The resulting light curves from July 27--28 are shown in Figure
\ref{fig:lcurve}.  On each night, we detected quasi-periodic infrared
flares with dereddened amplitudes of \mbox{$\sim 30$} mJy.  The flares
have durations of \mbox{$\sim$ 15--30} minutes and repeat on
timescales of \mbox{$\sim$ 30--60} minutes.  The accompanying X-ray
light curves show GRS~1915+105 undergoing a series of long ``dips''
and fast oscillations.  The particular pattern of X-ray variability
seen here has been observed many times in GRS~1915+105; it is the
``class $\alpha$'' state defined by \citet{Belloni2000}, characterized
by oscillations which grow longer and fainter with time before the
source enters an X-ray dip of duration \mbox{$\sim 1000$} seconds.

Our observations on July 29 were limited to \mbox{$\sim 1$} hour but
showed two infrared flares similar to those seen on July 27--28.  This
suggests that the state we observed lasted for more than two days.
The light curves from July 29 are not shown here because there were no
accompanying X-ray observations, but they are available upon request.
X-ray observations on July 30 indicated that GRS~1915+105 was no
longer undergoing oscillations and had entered a period of hard,
steady X-ray emission \citep[the ``class $\chi$'' state
of][]{Belloni2000}.

\subsection{Multiwavelength Features of the Light Curves}

As can be seen from Figure \ref{fig:lcurve}, each time the X-ray light
curve of GRS~1915+105 switches from a dip into a period of
oscillation, it is accompanied by an infrared flare.  Furthermore, the
flare appears to be triggered by a sharp X-ray ``spike'' of duration
several seconds which is present during each transition.

These features are broadly consistent with previous multiwavelength
observations of GRS~1915+105 during ``class B'' ejections
\citep[e.g.][]{Eiken98,Mirabel98}; in fact, many of the X-ray states
classified by \citet{Belloni2000} in which the source transitions
between a long dip and a period of oscillations are now known to
trigger radio (and therefore presumably infrared) flares
\citep{KleinWolt2002}.

What makes the current observations interesting is their high time
resolution and broad coverage of many individual episodes of jet
production, matched only by the similar observations of
\citet{Eiken98}.  By looking at how the subtle differences in X-ray
evolution between these two sets of observations lead to differences
in the infrared behavior, we can begin to unravel the details of the
complicated evolution of GRS~1915+105.

\begin{figure}
\epsscale{1.14} \plotone{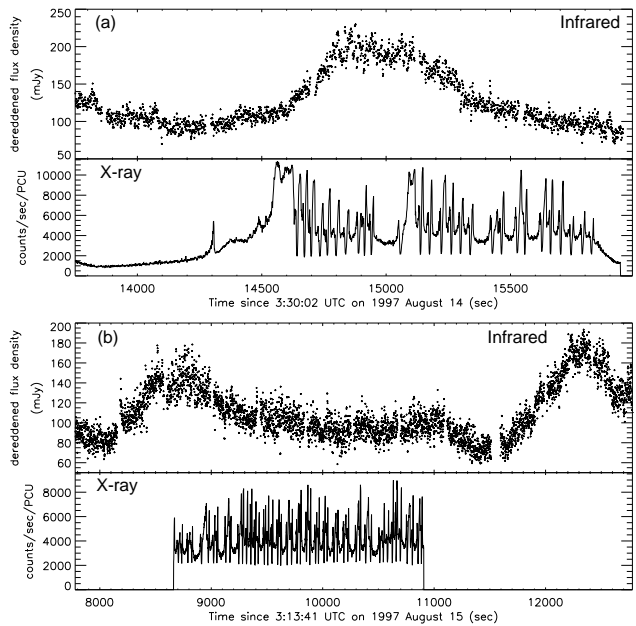} \epsscale{1}
\caption{Simultaneous infrared and X-ray light curves of GRS~1915+105
at 1-second resolution during August 1997, from data originally
presented in \citet{Eiken98}.  The infrared data have been dereddened
by 3.3 magnitudes before plotting.  In (a), the infrared flare appears
to end and the flux density decays back to its quiescent level while
the X-rays are still undergoing oscillations.  In (b), there is an
infrared flare beginning at \mbox{$\sim 8,000$} seconds which appears
to decouple from the X-rays (though there is no X-ray coverage for the
initial flare rise), but there is also a \mbox{$\sim 20$} mJy
``infrared excess'' that continues until \mbox{$\sim 11,000$} seconds
and that is associated with an unusually long series of X-ray
oscillations.}
\label{fig:1997}
\end{figure}

\subsection{Origin of the Infrared Flares} \label{origin}

Infrared flares similar to those presented here have been observed
many times from GRS~1915+105 and have been consistently interpreted as
synchrotron emission from a jet
\citep{Fender97,FenderPooley98,Eiken98,Mirabel98,Ueda}.  We propose
that the flares in Figure \ref{fig:lcurve} have the same origin.

This interpretation is supported by 15 GHz observations at the Ryle
Telescope which were obtained several hours before our July 28
observations (G. Pooley 2002, private communication).  Radio flares
with similar timescales as the infrared flares and amplitudes of
\mbox{$\sim$ 10--20} mJy (above a baseline flux density of \mbox{$\sim
10$} mJy) were observed.  If the infrared flares we observed had
similar radio counterparts, then the relatively flat infrared-to-radio
spectrum (as well as the \mbox{$\sim$ 10$^{9}$} to 10$^{10}$~K
brightness temperatures of the radio flares) indicates a nonthermal
origin.

\section{Infrared/X-Ray Correlation} \label{correlation}

The most interesting result from these observations may be the strong
correlation between the duration of each infrared flare and the
duration of its accompanying X-ray oscillation period.  All the flares
have similar rise times (\mbox{$\sim$ 200--300} seconds), but each
flare appears to ``wait'' to return to its quiescent level until the
X-rays stop oscillating.  This gives rise to asymmetric profiles for
some of the longer flares, particularly those on July 27.

How does this behavior compare to that seen in previous observations
of GRS~1915+105?  The flares observed in August 1997 by
\citet{Eiken98} were a few times stronger than those presented here,
with peak amplitudes as high as \mbox{$\sim 200$} mJy dereddened.  A
correlation between the infrared flare duration and the X-ray
oscillation period was not directly evident in 1997, and a few flares
appear to have decayed almost completely while the X-rays were still
undergoing strong oscillations.  An example of this behavior is shown
in Figure \ref{fig:1997}a.

However, there is some evidence for a smaller contribution to the
infrared light curve (on the \mbox{$\sim 20$} mJy level) associated
with the 1997 X-ray oscillations.  In particular, a long series of
oscillations occurred on August 15 at the same time as a period of
``infrared excess'' following a flare, in which the GRS~1915+105 flux
density remained steady at a level above its quiescent value (see
Figure \ref{fig:1997}b).  \citet{Eiken2000} showed that if each X-ray
oscillation in this series had a \mbox{$\sim$ 5--10} mJy infrared
flare associated with it, then the superposition of these flares could
reproduce the observed infrared excess.  Their conclusion was based on
the spectral similarity of the 1997 oscillations to isolated X-ray
oscillations in July 1998, each of which had an accompanying,
time-resolved \mbox{$\sim$ 5--10} mJy infrared flare (the ``class C''
flares).

\begin{figure}
\epsscale{1.1} \plotone{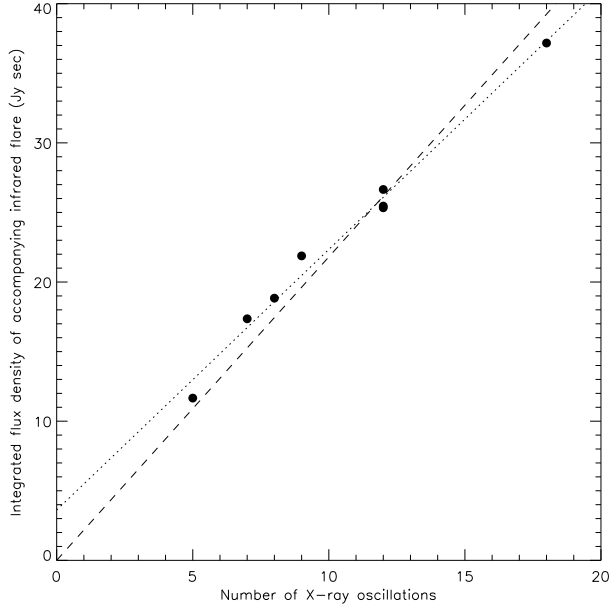} \epsscale{1}
\caption{The number of X-ray oscillations in each \mbox{$\sim 30$}
minute cycle plotted against the integrated flux density of the
infrared flare accompanying that cycle, from our observations on 2002
July 27--28.  Data from all eight infrared flares that had
uninterrupted X-ray coverage are shown in this figure, plotted as
points.  (Note that two of the points with 12 X-ray oscillations are
difficult to discern because they almost completely overlap.)  The
dotted line shows the best linear fit to the data, and the dashed line
shows the best linear fit constrained to pass through the origin.}
\label{fig:irxraycorr}
\end{figure}

The observations presented in our current work strengthen the argument
made by \citet{Eiken2000}.  The key piece of evidence comes from the
third infrared flare on July 27 (Figure \ref{fig:lcurve}a).  This
flare contains clear evidence for a ``subflare'' --- a secondary peak
that occurs after the flare has begun to decay.  Furthermore, the
subflare is associated with an isolated oscillation in the X-ray light
curve that occurs after an unusually long delay between it and the
previous set of oscillations.  It is therefore plausible that each
X-ray oscillation has a faint infrared subflare associated with it,
but in general the oscillations are so closely spaced in time that the
subflares blend together to create the illusion of one continuous
flare.\footnote{Note that there are a few other infrared flares in
Figure \ref{fig:lcurve} that appear to have very faint subflares, but
none of these events have good enough X-ray coverage to determine
whether they are associated with isolated X-ray oscillations.}

Unlike in the 1997 data, where these faint flares simply contribute an
``infrared excess'' on top of a larger episode of flaring behavior, we
propose that the \mbox{$\sim 30$} minute flares in 2002 are composed
{\it entirely} of these superimposed events.  This would immediately
explain the correlation between the infrared flare duration and the
X-ray oscillation period seen in 2002; the flare is only sustained by
the continual production of subflares associated with the X-ray
oscillations.

To test whether this idea is feasible, we show in Figure
\ref{fig:irxraycorr} the number of X-ray oscillations in each
\mbox{$\sim 30$} minute cycle plotted against the integrated flux
density of the infrared flare accompanying that cycle.  (The
integrated flux densities were calculated by smoothing the light curve
to 30 second resolution, determining the start and end times of each
flare by eye, subtracting out the baseline and integrating the result
numerically.)  It is clear from Figure \ref{fig:irxraycorr} that X-ray
cycles with more oscillations produce more infrared emission,
consistent with our suggestion.  Figure \ref{fig:irxraycorr} also
shows linear fits to the data; the best fit line has a slope of
\mbox{$\sim 1.9$} Jy~sec~oscillation$^{-1}$ and a vertical intercept
of \mbox{$\sim 3.6$} Jy~sec, while the best fit line constrained to
pass through the origin has a slope of \mbox{$\sim 2.2$}
Jy~sec~oscillation$^{-1}$.  Both lines appear to be reasonable fits,
and if we assume a plausible value of \mbox{$\sim 1.5$} Jy~sec for our
measurement errors in determining the integrated flux densities, then
both fits are found to be statistically acceptable, with reduced
chi-squared values of $\chi_{\nu}^{2}$ \mbox{$\sim 0.4$} and 1.1,
respectively.

\begin{figure}
\epsscale{1.1}
\plotone{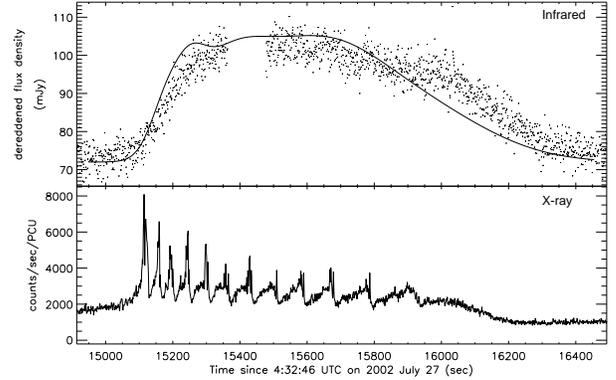}
\epsscale{1}
\caption{Simultaneous infrared and X-ray light curves of GRS~1915+105
at the time of the sixth infrared flare observed on 2002 July 27
(Figure \ref{fig:lcurve}a).  The infrared data (dereddened by 3.3
magnitudes) are plotted as points at 1-second resolution; the curve in
the top panel shows the results of our simulation (discussed in \S
\ref{correlation}), in which the infrared flare is modeled as a
superposition of small flares, one for each X-ray oscillation.}
\label{fig:simflare}
\end{figure}

\begin{figure*}
\epsscale{1.18} 
\plotone{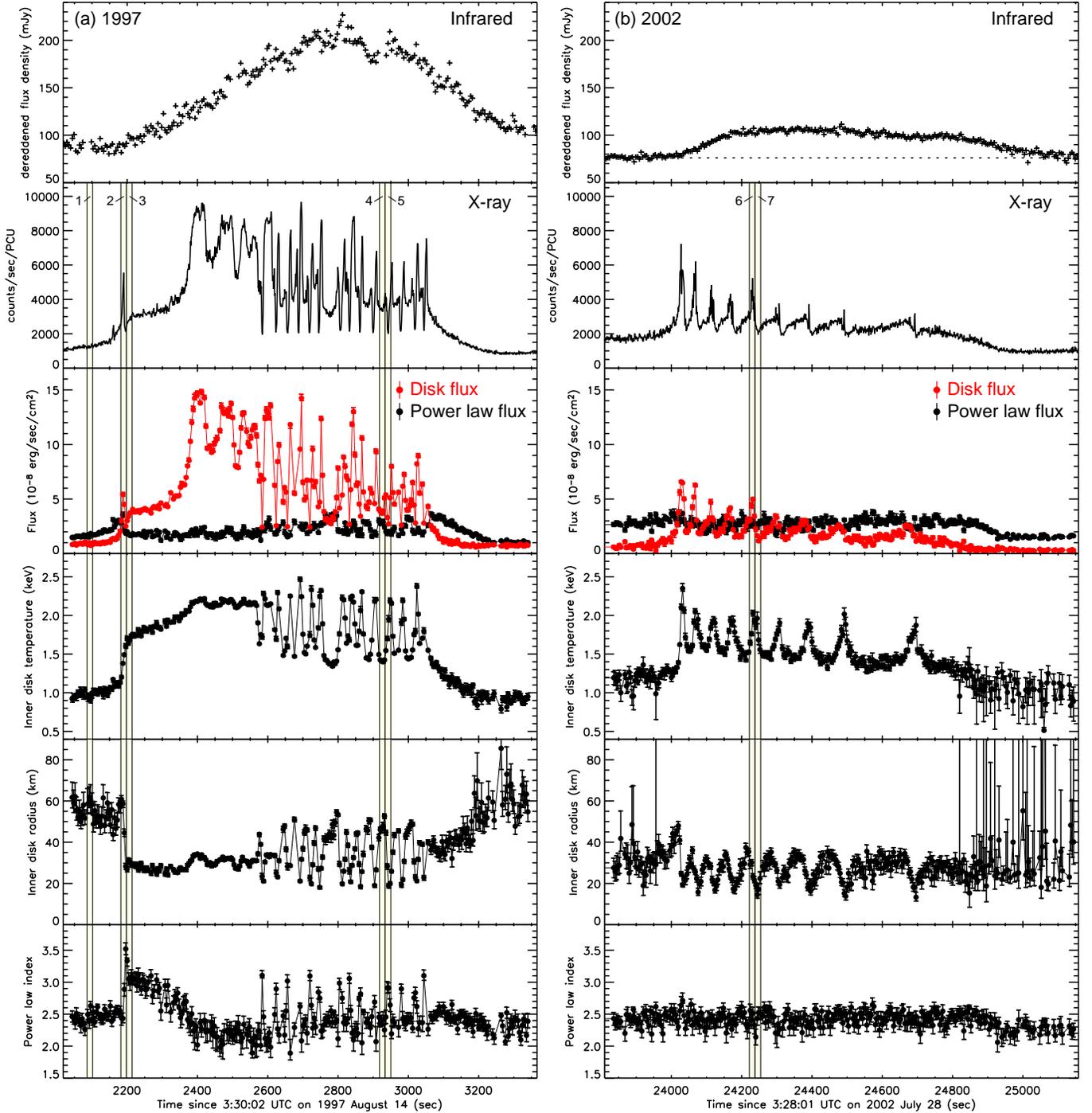} \epsscale{1}
\caption{Comparison of one cycle of jet formation in GRS~1915+105 from
observations in (a) 1997 and (b) 2002.  The top two panels in each
column show the infrared (dereddened by 3.3 magnitudes) and X-ray
light curves, respectively (at 1-second resolution, but the infrared
data have been smoothed to 5-second resolution in this plot); the
dotted horizontal line in (b) shows the baseline infrared level.  The
bottom four panels show the results of our X-ray spectral fitting at
4-second resolution, plotted along with \mbox{$\pm$ $1 \sigma$}
uncertainties; these include the unabsorbed \mbox{2--25} keV flux from
the multitemperature disk blackbody and power law components of the
spectrum, the temperature at the inner edge of the accretion disk, the
radius of the inner edge of the accretion disk, and the power law
index.  The vertical shaded regions numbered 1 through 7 show the
parts of the light curve from which the spectra plotted in Figure
\ref{fig:spectra} were extracted.}
\label{fig:specfit}
\end{figure*}

We also tried fitting a power law to the data in Figure
\ref{fig:irxraycorr} and found a slightly nonlinear relationship
(power law index \mbox{$\sim 0.8$}), but the quality of the fit was
not significantly better than that of the linear fit ($\chi_{\nu}^{2}$
\mbox{$\sim 0.34$}).  Therefore, the simplest conclusion we can draw
from our observations is that there is a roughly linear relationship
between the number of X-ray oscillations in each \mbox{$\sim 30$}
minute cycle and the strength of the infrared flare accompanying that
cycle.  In addition, we tested whether the infrared flare strength
might actually be related to the total amount of X-ray emission during
the oscillations (rather than the number of oscillations) and found no
evidence to support this; if the infrared flare strength is plotted
against the integrated X-ray flux in each cycle (above a baseline
level of 1800 counts/sec/PCU), the correlation has a larger scatter
than that shown in Figure \ref{fig:irxraycorr}, and both linear and
power law fits to the data are poor ($\chi_{\nu}^{2} > 2$).

The simplest interpretation of the linear relationship in Figure
\ref{fig:irxraycorr} is that each X-ray oscillation contributes a
constant amount of infrared emission to the overall light curve, in
the form of a faint flare.  In addition, the fact that the vertical
intercept of our fit is small (equivalent to an infrared excess of
only a few mJy over the typical duration of a flare) and possibly
consistent with zero supports our suggestion that the infrared events
are composed {\it entirely} of these superimposed faint flares.
Furthermore, the slope that we measure from our fit indicates that
each X-ray oscillation should be responsible for \mbox{$\sim 2$}
Jy~sec of infrared emission.  This number is slightly larger than, but
still consistent with, the integrated flux density in the subflare we
observed (between \mbox{$\sim$ 0.5} and 2 Jy~sec, depending on how we
choose to distinguish the ``subflare'' from the underlying main
flare).  It may also be reasonable to assume that an average X-ray
oscillation triggers a slightly larger subflare than the one we
observed, since this subflare was associated with one of the weakest
oscillations in its series.

To further test our idea, we attempted to simulate the infrared light
curve as a superposition of faint subflares.  We began our simulation
with a constant infrared background level of \mbox{$\sim 70$} mJy and
added one faint flare to this background for each oscillation in the
X-ray data.  We modeled each flare as a Gaussian with fixed amplitude
and full width at half maximum (FWHM).  Based on the observed
properties of the July 27 subflare, we chose the position of each
Gaussian so that its peak occurred 60 seconds after the peak of its
accompanying X-ray oscillation.  We performed several simulations with
different, fixed values of the amplitude and FWHM, since these
quantities were hard to determine from the observed subflare.

In general, these simulations did not fit the data well; the
superimposed flares we produced reached peaks which were initially too
high or dropped to values which were too low by the end of the
\mbox{$\sim 30$} minute cycle.  We therefore decided to allow the FWHM
of the flares to vary based on the duration of each X-ray oscillation.
We chose a simple linear scaling, in which the FWHM of each flare was
taken to be proportional to the rise time of its accompanying X-ray
oscillation.  Results of this simulation for the sixth infrared flare
on July 27 are shown in Figure \ref{fig:simflare}; in this case, we
fixed the amplitude of each Gaussian to be 8 mJy and set each
Gaussian's FWHM to be a factor of 5 greater than the rise time of its
accompanying X-ray oscillation.  This choice guaranteed that a typical
oscillation in each X-ray series, with a rise time of \mbox{$\sim 50$}
seconds, had an associated infrared flare with a FWHM of 250 seconds
and an integrated flux density of \mbox{$\sim 2$} Jy~sec, consistent
with the slope of the fit in Figure
\ref{fig:irxraycorr}.\footnote{Note that by allowing the FWHM to vary,
we are also varying the amount of infrared emission associated with
each X-ray oscillation.  However, this does not contradict the linear
relationship found in Figure \ref{fig:irxraycorr}; it turns out that
each X-ray cycle has about an equal number of oscillations with rise
times above and below the typical value of \mbox{$\sim 50$} seconds,
so the stronger infrared flares may ``cancel out'' the weaker ones and
still lead to a roughly linear relationship.}

Simulations such as the one in Figure \ref{fig:simflare} do a good job
of reproducing the observations, especially given the gross
approximations we are making.  Even so, some difficulties are
encountered.  The July 27 subflare is not reproduced using the above
procedure; its accompanying X-ray oscillation has an extremely long
rise time, and therefore the simulation tries to model it with an
unrealistically large FWHM.  In general, the simulations work better
at the beginning of each \mbox{$\sim 30$} minute cycle than at the
end, suggesting that there is residual infrared emission associated
with the end of each X-ray cycle that is not captured in our model.
Nonetheless, the results shown in Figure \ref{fig:simflare} are fairly
typical, and we can be confident that the 2002 flares consist
primarily of superimposed faint flares associated with each X-ray
oscillation.

\begin{figure*}
\epsscale{1.18}
\plotone{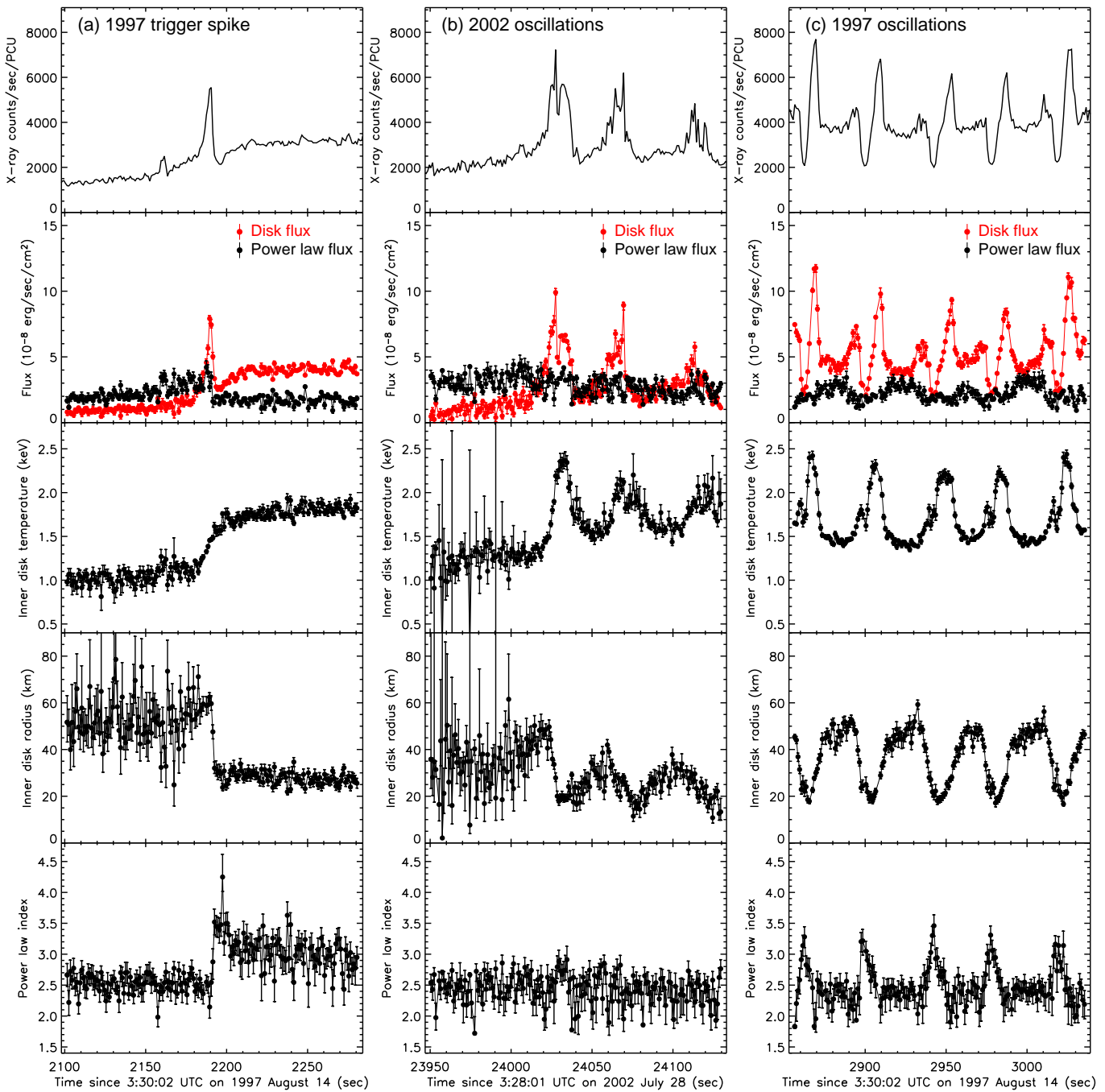}
\epsscale{1}
\caption{Comparison of the X-ray spectral evolution of GRS~1915+105
during (a) the 1997 trigger spike (from Figure \ref{fig:specfit}a),
(b) the initial spike in 2002 and the oscillations which immediately
follow it (Figure \ref{fig:specfit}b), and (c) a typical set of
oscillations in 1997 (Figure \ref{fig:specfit}a).  The top panel in
each column shows the X-ray light curve (at 1-second resolution), and
the bottom four panels show the same X-ray spectral fitting parameters
plotted in Figure \ref{fig:specfit}, but the fits shown here were
performed at 1-second resolution.  It is clear from this figure that
the trigger spike in (a) is spectrally different from the other
oscillations in 1997 and 2002.}
\label{fig:1sec}
\end{figure*}

\section{X-Ray Spectral Evolution} \label{specevolve}

We can compare our X-ray light curves and spectra to those obtained in
August 1997 by \citet{Eiken98}.  Though both sets of observations show
transitions between X-ray dips and oscillations, the details of the
behavior are different; in the classification of \citet{Belloni2000},
GRS~1915+105 was in the ``class $\beta$'' state in 1997 and the
``class $\alpha$'' state in 2002.

In Figure \ref{fig:specfit}, we present one cycle of jet formation
from each of the 1997 and 2002 observations.  Several parameters
obtained from our spectral fitting are plotted at 4-second resolution:
the unabsorbed \mbox{2--25} keV flux from the multitemperature disk
blackbody and power law components of the spectrum, the temperature at
the inner edge of the accretion disk $T_{in}$, the radius of the inner
edge $R_{in}$, and the power law index $\gamma$ (where the flux
density due to the power law, at photon energy $E$, is $\propto
E^{-\gamma}$).  The fits shown in Figure \ref{fig:specfit} have
typical reduced chi-squared values of $\chi_{\nu}^{2}$ \mbox{$\sim 1.
2$}, and all fits with $\chi_{\nu}^{2} > 2$, as well as those for
which XSPEC was unable to return uncertainties on the fit parameters,
were excluded from the plot.  Note that $R_{in}$ is calculated from
the disk blackbody normalization ($R_{in} \propto \sqrt{N}$, where $N$
is the normalization) assuming a distance to GRS~1915+105 of 11 kpc
and an accretion disk inclined $66\degr$ to the line of sight
\citep{Fender99}, but the values derived from this procedure should
not be viewed as accurate estimates of the physical size of the disk
\citep[see \S \ref{sec:thermspike} and][]{Merloni2000}.

\subsection{1997 and 2002 State Changes}

Figure \ref{fig:specfit} shows that in both 1997 and 2002, the X-ray
spike which appears near the beginning of the infrared flare (at
\mbox{$\sim 2,200$} seconds in 1997 and \mbox{$\sim 24,000$} seconds
in 2002)  represents a ``state change'' within the accretion disk, in
the sense that it initiates changes that persist after the spike is
finished.  During the X-ray dip, the source is in a spectrally hard
state in which the disk blackbody component is nearly absent, but the
increase in $T_{in}$ and in the disk emission which occur at the spike
indicate that the disk becomes hot and visible at this point and
remains that way throughout the X-ray oscillations.

In 1997, there is a prolonged period of a few hundred seconds after
the X-ray spike when the inner accretion disk has ``turned on'' but
before the period of fast oscillations.  It is clear that the infrared
flare begins well before the oscillations, so although we may be able
to ascribe some of the infrared emission to blended together
``subflares'' associated with each oscillation (as discussed in \S
\ref{correlation}), there must be some other mechanism at work which
leads to the initial rise of the infrared light curve right around the
moment of the spike and allows it to reach a peak amplitude of
\mbox{$\sim 100$} mJy.

In 2002, however, the spike seems to represent the first oscillation
in a series; it is spectrally similar to the oscillations which
immediately follow it, and differs mainly in that it is the strongest
and fastest oscillation in the group.  This is consistent with the
idea that the flares observed in 2002 consist entirely of superimposed
\mbox{$\sim$ 5--10} mJy subflares associated with each oscillation,
and that no abnormally large ejection occurs at the moment of the
first spike.

The differences between the 1997 and 2002 behavior can be more clearly
seen in Figure \ref{fig:1sec}, where we have plotted the results of
X-ray spectral fitting performed at high time resolution (1 second).
Again, all fits with $\chi_{\nu}^{2} > 2$, as well as those for which
XSPEC was unable to return uncertainties on the fit parameters, were
excluded from the plot.  Figure \ref{fig:1sec}a shows the initial
X-ray spike in 1997, Figure \ref{fig:1sec}b shows the initial X-ray
spike in 2002 and the oscillations which immediately follow it, and
Figure \ref{fig:1sec}c shows a typical set of oscillations in 1997.

It is clear from Figure \ref{fig:1sec} that although the details of
the 1997 and 2002 oscillations are different, there are also many
similarities:  both consist primarily of periodic changes in the disk
emission and $T_{in}$.  The initial oscillation in 2002, which signals
the beginning of the disk activity and coincides with the rise of the
infrared flare, is no different in this regard.  Figure
\ref{fig:1sec}a, however, shows that the initial spike in 1997 (which
we refer to as the ``trigger'' spike) is an entirely different
phenomenon.  During this spike, $T_{in}$ rises monotonically to
\mbox{$\sim 1.8$} keV, where it remains, while the power law component
of the spectrum does not vary noticeably during the spike but changes
sharply as the spike ends.  Similar behavior was observed by
\citet{migliari}, but our spectral fits are performed at higher time
resolution and therefore confirm that the power law does not change
significantly even during the last few seconds of the spike's rise.

\subsection{1997 Trigger Spike}

Given the importance of the 1997 trigger spike and its possible
connection to large ejections in GRS~1915+105, it is useful to examine
its evolution in a model-independent way.  This is shown in Figure
\ref{fig:spectra}a, where we plot three X-ray spectra associated with
the trigger spike (the parts of the X-ray light curve from which these
spectra have been extracted are indicated by the vertical shaded
regions numbered 1 through 3 in Figure \ref{fig:specfit}a).  We
extracted the spectra at a time resolution of 16 seconds using PCA
{\tt Standard-2} data, which have higher energy resolution than the
Binned and Event mode data used for the fitting procedure.

As can be seen from Figure \ref{fig:spectra}a, between regions  1 and
2 (as the spike begins) there is virtually no change in the hard
(\mbox{$\ga 20$} keV) X-ray emission, but the soft X-rays brighten
dramatically.  After the drop at the end of the spike, however (region
3), the spectrum steepens sharply, and the hard X-rays become nearly
undetectable.  The soft X-rays in region 3 retain a similar spectral
shape as in region 2, but decrease in intensity.

In this model-independent description, the trigger spike consists of:
\begin{enumerate}
\item A factor of a few increase in the soft X-ray emission on a
timescale of \mbox{$\sim 20$} seconds or longer.
\item A dramatic, factor of a few decrease in the hard X-ray emission
and less dramatic decrease in the soft X-ray emission, on a timescale
of a few seconds.
\end{enumerate}

How do these changes in the soft and hard X-ray emission during the
trigger spike relate to each other physically?  There are several
possible ways to understand this, depending on how much we believe the
accuracy of the multitemperature disk blackbody model that we have
employed above.

\subsubsection{Nonthermal Models for the Spike} \label{sec:nonthermalspike}

Although our spectral fitting indicates that the soft X-ray excess
during the trigger spike consists of thermal emission from the
accretion disk, this is not an ironclad conclusion.  The energy
resolution of our data is quite poor, and it is possible that other
models could reproduce this brief burst of soft X-ray emission.
Before the spike, the accretion disk component of the spectrum is
extremely weak, and in fact, we are able to obtain acceptable fits at
1-second resolution ($\chi_{\nu}^{2}$ \mbox{$\sim 1.4$}) for a model
that only includes a power law and no accretion disk.  Thus, it is
possible to conclude that the spike is simply a ``blip'' on an
otherwise smooth rise of the disk flux that corresponds (along with
the smooth rise in $T_{in}$) to the accretion disk ``turning on.''  In
this picture, when the 1997 X-ray light curve transitions from a dip
into a period of oscillations, the thermal emission from the disk {\it
gradually} becomes visible (as it does in 2002).  At some point during
this transition, however, a \mbox{$\sim 10$} second nonthermal burst
of soft X-rays is released, creating the spike.  This interpretation
certainly seems plausible in light of the second panel of Figure
\ref{fig:1sec}a, which shows the disk flux rising smoothly before the
spike and finishing its smooth rise afterwards; the spike can easily
be imagined as a sharp feature superimposed on an otherwise smooth
event.

\begin{figure}
\epsscale{1.15} 
\plotone{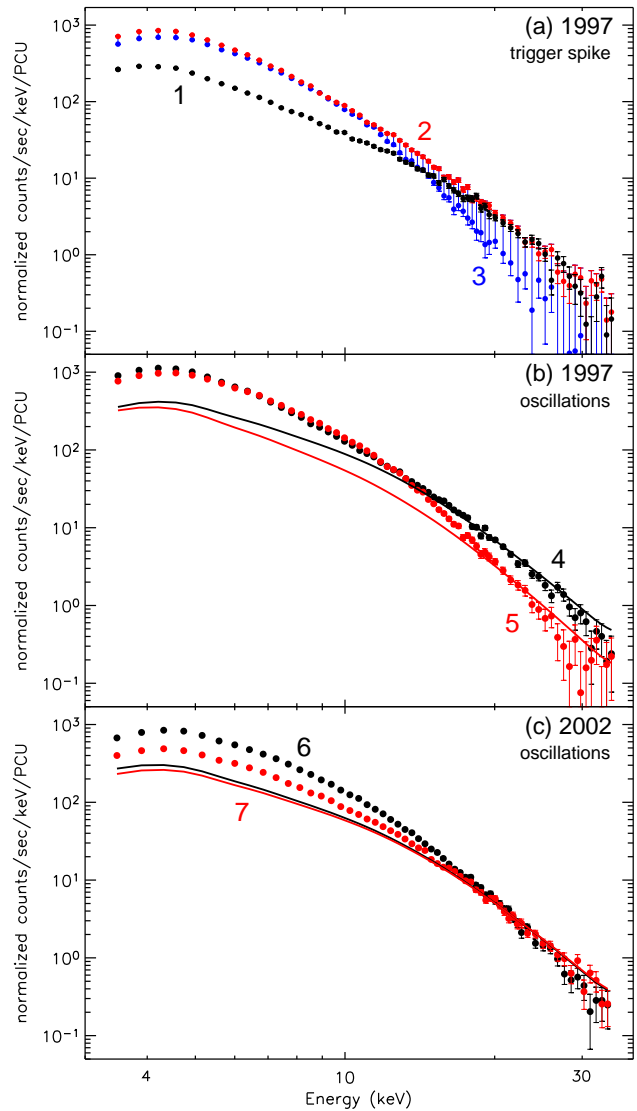} \epsscale{1}
\caption{X-ray spectra of GRS~1915+105 during (a) the 1997 trigger
spike in Figure \ref{fig:specfit}a, (b) the 1997 oscillations in
Figure \ref{fig:specfit}a, and (c) the 2002 oscillations in Figure
\ref{fig:specfit}b.  The spectra are numbered 1 through 7,
corresponding to the regions of the X-ray light curve from which they
were extracted (see Figure \ref{fig:specfit}).  The spectra were
obtained at 16-second resolution and are plotted as points with
\mbox{$\pm$ $1 \sigma$} uncertainties; the curves in (b) and (c) show
the power law component of the fit to each spectrum, modified by
hydrogen absorption (this component accounts for all of the hard X-ray
emission and a fraction of the soft X-ray emission).}
\label{fig:spectra}
\end{figure}

If this is the case, then when the nonthermal soft X-ray emission
disappears at the end of the spike, the nonthermal hard X-ray emission
(from the power law component of the spectrum) is also observed to
drop (Figure \ref{fig:spectra}a).  The origin of the power law in
X-ray binaries is a subject of some debate; it is generally assumed to
represent either inverse Compton emission from a corona of
relativistic electrons \citep{Poutanen} which may constitute the base
of a jet \citep{FenderCompton}, or the high frequency tail of
synchrotron emission from the jet itself \citep{Markoff,Markoff2003}.
In either case, it comes from a region close to the accretion disk, so
the decrease in the power law emission at the end of the spike
indicates that changes are occurring immediately outside of the
accretion disk at this time.  Meanwhile, the infrared flare also
begins, as energy is dissipated in the jet further downstream from the
disk.

To test the nonthermal scenario, we tried fitting the soft X-ray
emission during the spike as a power law (that is, we fit two power
laws to the spectrum, the original one which extended to the hard
X-rays, and a stronger, steeper one which covered the soft X-ray
excess) but were not able to obtain acceptable fits at 1-second
resolution.  This implies the need for a ``curved'' spectral component
to explain the emission during the spike; however, there are many
possible sources for such a component besides thermal emission from
the disk.

The above suggestions are tentative, and much more data would be
required to confirm them.  Nevertheless, a logical picture can be
constructed in which the spike does not originate from thermal
emission in the accretion disk.  This picture would be consistent with
several recent models, including the ``magnetic bomb''
\citep{VanPutten} and ``magnetic flood'' \citep{Tagger} models for
GRS~1915+105 and the more general model proposed by \citet{Livio}.
These models all evoke the existence of large scale magnetic fields
during the X-ray dip and associate the state transition at the end of
the dip with the destruction of these magnetic fields, perhaps through
a reconnection event that produces a \mbox{$\sim 10$} second burst of
nonthermal emission along with an ejection of material.

It is worth pointing out that our 2002 observations strengthen some of
the above models, since they show that the large ejections
(\mbox{$\sim 100$} mJy infrared flares which decouple from the X-ray
light curve) only seem to occur in the presence of the trigger spike.
However, they also show that the \mbox{$\sim 30$} minute state
transitions can occur {\it without} an initial large infrared event.
Why is this the case?  If the spike and ejection represent the
destruction of the magnetic field which triggers the state transition,
then why do state transitions sometimes seem to require a trigger
spike and ejection \mbox{$\sim 300$} seconds before the fast
oscillations, but other times occur without them and jump immediately
into the oscillations?  Ideally, a model for the GRS~1915+105 behavior
should answer these questions.

\subsubsection{Thermal Accretion Disk Models for the Spike} \label{sec:thermspike}

If we instead elect to believe that the soft X-ray excess during the
trigger spike is due to thermal emission from the accretion disk, we
need to explain how the disk interacts with the corona or jet (which
is responsible for the power law component of the spectrum) in order
to produce the observed emission.

\citet{yadav} suggested a general picture for the spike in which the
soft X-ray luminosity increases until it reaches a critical value,
thus triggering the ejection and subsequent changes in the light curve
and spectrum.  In this picture, all of the ``action'' takes place at
the end of the spike; at this time, the X-ray power law and infrared
light curve change in a way that is consistent with jet ejection (just
as they did in the nonthermal spike scenario of \S
\ref{sec:nonthermalspike}), but this change is now accompanied by a
sharp drop in the accretion disk luminosity.  This scenario has some
advantages for observations in which the trigger spike does not have a
well-defined beginning but rather seems to rise gradually to its
maximum throughout the preceding X-ray dip \citep[see, for example,
the observation presented in][]{migliari}.

What could cause the accretion disk luminosity to decrease at the end
of the spike, at the same time as the power law component changes?
One possibility is that the drop in disk luminosity indicates a sharp
increase in the fraction of the accretion power which is being
dissipated in the GRS~1915+105 jet.  Although detailed modeling of the
disk spectrum would be required to confirm this suggestion, it is
consistent with the changes observed in the disk parameters at the end
of the spike (as discussed below), and it is also consistent with the
observed changes in the X-ray power law and infrared light curve,
which show that material may be ejected into the jet at the end of the
spike.

To study in detail the changes in the disk parameters at the end of
the spike, it is useful to examine Figure \ref{fig:1sec}a.  As can be
seen from this figure, it is only the flux of radiation coming from
the accretion disk which changes dramatically at the end of the spike,
not the basic shape of the disk spectrum; the end of the spike occurs
because of a sharp drop in $R_{in}$ over a few seconds, while $T_{in}$
increases over a slower timescale (\mbox{$\sim 20$} seconds)
throughout the spike.

This behavior is puzzling.  In general, changes in the disk parameters
occur on similar timescales, in which case an increase in $T_{in}$
coupled with a decrease in $R_{in}$ can be understood by a model in
which the physical conditions in the outer part of the disk remain
similar but the inner part of the disk has ``turned on,'' either by
refilling with matter or by beginning to emit detectable radiation
\citep{Belloni97a,Belloni97b}.  This scenario has been successfully
simulated by \citet{Watarai2003}; the decrease in $R_{in}$ is due to
the fact that at high accretion rates \citetext{where the ``slim
disk'' model of \citealp{Abramowicz} applies}, significant thermal
radiation can emerge from inside the last stable circular orbit around
the black hole \citep{Watarai2000}.  However, the fact that in our
observations the drop in $R_{in}$ occurs {\it rapidly} at the end of
the spike, without any corresponding rapid changes in $T_{in}$, is
more difficult to understand, and it is therefore necessary to look
for other interpretations for what is happening to the radiation from
the accretion disk at the end of the spike.

\citet{Rodriguez} point out that when the normalization of the
multitemperature disk blackbody is observed to drop, the simplest
interpretation is that the effective area of the emitting region of
the disk has decreased; the calculation of $R_{in}$ from the
normalization is model-dependent and not necessarily reliable.  A drop
in the fitted value of $R_{in}$ can therefore be interpreted as the
appearance of a hot spot or spiral shock pattern in the disk
\citep{Rodriguez}, or more generally as a situation where some
representative portion of the disk stops emitting radiation which we
receive in the soft X-ray band \citep[see also][]{Muno2001}.  In this
vein, \citet{Merloni2000} showed that if a more realistic model of the
disk flow is used to simulate the X-ray spectrum, then changes in
several parameters, including the fraction of the accretion power
dissipated in regions outside of the disk (for example, in a corona or
jet), could lead to apparent changes of $R_{in}$ by a factor of
\mbox{$\sim 2$} when the multitemperature disk blackbody model is used
to fit the spectrum.

It is therefore plausible that the end of the spike coincides with a
transfer of accretion power from the disk (where it is dissipated
thermally) to the jet (where it initiates changes in the X-ray power
law and infrared flare).  During the \mbox{$\sim 300$} second break
between the spike and the subsequent oscillations, the disk then
settles into the standard ``slim disk'' state, where the oscillations
can occur.  However, as in \S \ref{sec:nonthermalspike}, it is still
not clear from this scenario what physical parameter triggers the
sharp transition and jet ejection in our 1997 observations but not in
2002.

\subsubsection{The Spike in the Context of Other X-ray Binaries} \label{sec:otherspike}

\citet{FenderGallo} have proposed a unified model for jet formation in
X-ray binary systems.  The model is ``unified'' in the sense that the
trigger spike in GRS~1915+105 is shown to have counterparts in several
other sources, where a radio flare is produced around the time of a
peak in the X-ray light curve that occurs during the transition from a
hard to soft spectral state.  \citet{FenderGallo} proposed a physical
model in which a steady jet exists during the spectrally hard X-ray
state preceding the spike and the speed of the jet increases sharply
as the X-ray spectrum becomes softer; the jet is then quenched
completely once the softness reaches a critical value.  Thus,
synchrotron flares are produced via internal shocks in the preexisting
jet around the time of the X-ray spike, when the jet velocity is
increasing the fastest.

Our observations are relevant to this model in several ways.  First,
we have shown that the sharp X-ray trigger spike observed in 1997 is
associated with the appearance of large infrared flares that do not
occur in 2002, when the spike is absent.  This would seem to be
evidence in favor of the \citet{FenderGallo} model.  On the other
hand, a hard-to-soft state transition clearly occurs in 2002 as well,
and the first oscillation in each 2002 cycle might therefore meet the
definition of a trigger spike in this sense (see also \S
\ref{continuum}).  If so, the model would need to explain why the 1997
trigger spike is associated with a \mbox{$\sim 100$} mJy infrared
flare, while the 2002 spike is only associated with a \mbox{$\sim$
5--10} mJy flare that is rivaled, at least in the infrared, by the
flares that immediately follow it.

There are several reasons why the strength of the flare might vary.
As suggested by \citet{Vadawale}, the flare strength may be related to
the total amount of material in the path of the faster jet --- that
is, if more material was ejected during the spectrally hard X-ray dip
that preceded the spike, the resulting flare might be stronger.
However, our observations do not show any particular relationship
between the duration of each X-ray dip and the strength of the
subsequent infrared flare.  Therefore, there is no evidence to suggest
that this effect is responsible for the stronger flares in 1997 as
compared to 2002.

Another mechanism for varying the strength of the flare in the
\citet{FenderGallo} model is to change the speed of the jet.
\citet{FenderBelloni} suggested that the proposed increase in jet
speed as the X-ray spectrum softens may be due to the inner edge of
the accretion disk moving closer to the black hole, where the escape
velocity increases rapidly.  If we take the results of our X-ray
spectral fits in Figure \ref{fig:1sec} at face value, we can attempt
to test this suggestion, by seeing whether there are significant
differences in the evolution of $R_{in}$ between 1997 and 2002.

It can be seen from Figure \ref{fig:1sec} that although $R_{in}$
decreases during the spike, it does so by a similar order of magnitude
in 1997 as in 2002, suggesting that there may not be any large
differences in the jet velocity.  On the other hand, the value of
$R_{in}$ just before the spike appears to be slightly larger in 1997
than it is in 2002.  Therefore, if the argument of
\citet{FenderBelloni} is correct, the velocity of the steady jet is
slower in 1997, leading to a greater velocity differential and
stronger internal shock when the highly relativistic jet is released
during the spike.  In addition, the decrease in $R_{in}$ happens on a
faster timescale in 1997 than in 2002, which could mean that the jet
velocity increases faster, leading to an internal shock in 1997 which
occurs closer to the disk and therefore in the peak infrared-emitting
region of the jet.

However, this interpretation carries with it many caveats since, as we
have pointed out in \S \ref{sec:thermspike}, the observed change in
$R_{in}$ may not correspond to an actual change in the inner radius of
the disk, and even if it does, it only corresponds to the radius of
the {\it thermally emitting} part of the disk (i.e. the region that is
optically thick), which is not necessarily the same as the region from
which the jet is ejected.  Furthermore, there is a considerable range
of X-ray behavior in both the 1997 and 2002 observations, so any
conclusions that apply to the data shown in Figure \ref{fig:1sec} do
not necessarily apply as strongly to the ``class $\alpha$'' and
``class $\beta$'' states as a whole.

Another way in which our observations are relevant to the
\citet{FenderGallo} model concerns the infrared subflare and our
suggestion that each X-ray oscillation in 2002 is associated with a
similar amount of infrared emission (see \S \ref{correlation}).  If
these faint infrared flares are interpreted as emission from the jet,
then the jet in GRS~1915+105 may be active during the time of the soft
X-ray oscillations, in contrast to the suggestion of
\citet{FenderGallo}.  On the other hand, we cannot conclusively prove
that the subflares represent jet emission (see \S \ref{sec:classc}),
and even if they do, it is not clear exactly where the ``dividing
line'' between jet and non-jet states in the \citet{FenderGallo} model
actually exists; it is possible that the low points of the 2002 X-ray
oscillations (where \mbox{$T_{in} \lesssim 1.5$} keV) are spectrally
hard enough to produce a steady jet.

In general, our observations do not contradict the \citet{FenderGallo}
model at this stage of its development; they will, however, likely be
important in testing a more quantitative model based on the ideas that
\citet{FenderGallo} proposed.  Furthermore, any successful model will
need to explain what parameter triggers the sharp X-ray state
transitions and infrared-bright ejections that are observed in 1997
but not in 2002.

\begin{figure*}
\epsscale{1.07}
\plotone{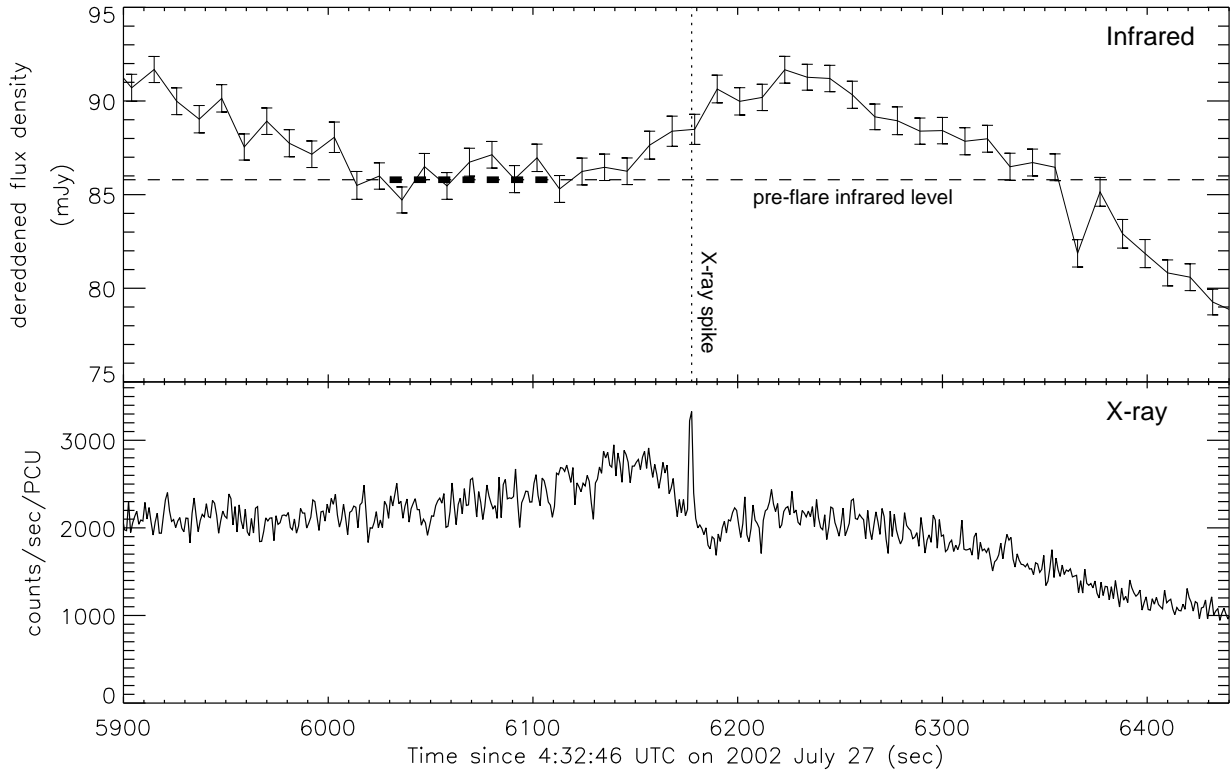}
\epsscale{1}
\caption{Simultaneous infrared and X-ray light curves of GRS~1915+105
at the time of the infrared ``subflare'' observed on 2002 July 27
(Figure \ref{fig:lcurve}a).  The X-ray data are at 1-second
resolution, and the infrared data (dereddened by 3.3 magnitudes) have
been smoothed to 11-second resolution to increase the signal-to-noise
ratio; they are plotted along with their \mbox{$\pm$ $1 \sigma$}
uncertainties.  The dotted vertical line marks the time of the X-ray
spike, and the dashed horizontal line marks the median value of the
infrared light curve calculated between 6,030 and 6,110 seconds (the
bold region of the dashed line).  The subflare appears to rise above
this value \mbox{$\sim 30$} seconds before the X-ray spike begins.}
\label{fig:subflare}
\end{figure*}

\subsection{1997 and 2002 Oscillations} \label{oscillations}

Perhaps the most interesting difference between the 1997 and 2002
X-ray spectra is in the evolution of the power law component.  In
1997, the power law index undergoes much more dramatic changes than in
2002, especially during the period of oscillations (see Figures
\ref{fig:specfit} and \ref{fig:1sec}).  If this difference is real, it
would suggest that significant changes are occurring in the corona or
jet during the 1997 oscillations that do not take place during 2002,
which is somewhat puzzling since both sets of oscillations produce
similar ``class C'' infrared flares (see \S \ref{correlation}).

We investigated the possibility that the observed variations in the
power law index were artificial.  For example, if the multitemperature
disk blackbody model provided a poor fit to the soft X-ray data, this
could force the fitting routine to alter the power law component in
order to fix the problems with the disk model on the soft X-ray end.
Since the power law contributes significantly to both the soft {\it
and} hard X-rays, the model spectrum might then fit the hard X-rays
poorly, which would not necessarily be noticed in our fitting routine
because the hard X-rays only provide a small percentage of the total
flux and contribute little to the chi-squared value of the fit.
Therefore, it might be possible to obtain changes in the fitted power
law index even if there were no changes in the actual flux of hard
X-ray photons, which comprise the bulk of the power law shape.

To get a handle on whether or not this is occurring, we show plots of
X-ray spectra from the 1997 and 2002 oscillations in Figure
\ref{fig:spectra}b and \ref{fig:spectra}c.  The parts of the X-ray
light curve from which these spectra have been extracted are indicated
by the vertical shaded regions numbered 4 through 7 in Figure
\ref{fig:specfit}.  As before, we extracted the spectra at a time
resolution of 16 seconds using PCA {\tt Standard-2} data, in order to
obtain the highest possible energy resolution.

It is clear from Figure \ref{fig:spectra} that the spectrum at
energies \mbox{$\ga 20$} keV does not change significantly during the
2002 oscillations but does during 1997, where the hard X-ray flux
appears to drop by a factor of a few as the light curve enters the low
point of its oscillation \citep[the low point of the oscillation is an
example of an X-ray ``soft dip'' discussed in e.g.][]{Eiken2000}.
Furthermore, the solid lines in Figure \ref{fig:spectra} show the
power law fit to the data; at high energies, the fit does seem to
match the spectrum relatively well.  There is a tentative suggestion
that the actual data is steeper than the fitted power law, but it
appears to be a minor effect.

We attempted to investigate this issue further by performing power law
fits to the hard X-ray data only, thereby circumventing any problems
introduced by the multitemperature disk blackbody model.  We found
that the power law index did appear to systematically increase by a
value of \mbox{$\sim 0.3$} to 1 when this procedure was used, but we
were unable to obtain high enough signal-to-noise to reliably
investigate the variability properties.  Therefore, while we can
conclusively state that the high energy power law component changes
much more dramatically in 1997 than 2002, our data showing that this
change specifically occurs in the power law index are less robust.

Finally, although the power law is much more variable in 1997 than
2002, there does appear to be a small peak in the fitted power law
index during the first oscillation in 2002 (Figure \ref{fig:1sec}b).
There is also tentative evidence for this behavior in the first
oscillation of other X-ray cycles in our 2002 observations.  These
detections are very close to our signal-to-noise limit, and thus we
cannot determine whether the power law component varies in {\it all}
the 2002 oscillations but is in general too weak for us to detect
(since the subsequent oscillations are not as strong as the first
one), or whether the first X-ray oscillation is in fact physically
distinct, and produces power law variations even though the others do
not.

\section{Origin of the ``Class C'' Subflares} \label{sec:classc}

It is interesting to speculate on the origin of the \mbox{$\sim$
5--10} mJy ``class C'' infrared flares which seem so ubiquitous in
GRS~1915+105.

\citet{Eiken2000} showed that when these flares were observed in July
1998, each flare {\it preceded} its associated X-ray oscillation by
\mbox{$\sim$ 200--600} seconds, including examples in which the
infrared light curve peaked and started to decay before the X-ray
disruption began.  To see if this property applies to the events we
observed in 2002, we show in Figure \ref{fig:subflare} a closer view
of the isolated ``subflare'' from our first night of observations.
The X-ray oscillation associated with this subflare consists of a
smooth rise in the accretion disk flux, followed by a sharp spike.  As
can be seen from Figure \ref{fig:subflare}, the subflare peaks after
the X-ray spike, but its initial rise does seem to begin \mbox{$\sim
30$} seconds before the spike.  This evidence is tenuous, but it is
another hint that the \mbox{$\sim$ 5--10} mJy flares in GRS~1915+105
might all have a similar cause.

Note that the X-ray event shown in Figure \ref{fig:subflare} is in no
way unusual except for the long delay between it and the previous
oscillations; a similar spike is seen at the end of most, if not all,
of the 2002 oscillations (for example, see the last event in Figure
\ref{fig:1sec}b).  Similar oscillations have been discussed by
\citet{taam}, where the spike is referred to as a ``secondary burst.''
Spectral fitting of these spikes is inconclusive but indicates that
they may consist of a momentary increase in the flux from the
accretion disk {\it and} power law components of the spectrum, which
makes them different from any other part of the 1997 or 2002
oscillations.  Note that the ``class C'' infrared flares originally
defined by \citet{Eiken2000} were associated with ``soft dips'' in the
X-ray light curve that occur in the 1997 oscillations (see \S
\ref{oscillations}), but for the purposes of this paper, we will refer
to any infrared subflare that occurs during the \mbox{$\sim 30$}
minute events and is associated with an individual oscillation as a
``class C'' flare.

We cannot directly rule out a thermal origin for the 2002 subflare
(since it is an isolated event with no radio coverage), but the
near-simultaneous presence of radio flares during our observations
(see \S \ref{origin}), combined with the idea that all of the infrared
flaring behavior in 2002 was due to superimposed smaller events (see
\S \ref{correlation}), strongly suggests that each small infrared
flare has a radio counterpart, and therefore a nonthermal origin (see,
however, \S \ref{irradioflares}).  In addition, \citet{Eiken2000}
argued that the July 1998 ``class C'' infrared flares were probably
related to synchrotron jet emission, based on observations by
\citet{Feroci} of a delayed, \mbox{$\sim 40$} mJy radio counterpart to
an X-ray oscillation spectrally similar to the July 1998 oscillations
(again, however, we point out that the 2002 oscillations do not share
this similarity).

It is clear that the ``class C'' flares are a complex phenomenon, with
a variety of X-ray counterparts.  There is also further fast
variability within the ``class B'' events that has not been discussed
here: large (up to \mbox{$\sim 70$} mJy) subflares in some of the 1997
data that unfortunately have no X-ray coverage, and smaller
(\mbox{$\sim$ 10--20} mJy) flares that do {\it not} occur during the
X-ray oscillations but rather during the initial rise of the infrared
light curve, right after the trigger spike \citep{Rothstein}.  If we
accept a synchrotron jet origin for all the flares in GRS~1915+105,
then the observations discussed here show that the accretion disk can
interact with the jet under a wide variety of timescales and
circumstances.

What is the nature of this interaction?  The fact that at least some
of the flares precede their associated X-ray oscillation argues for an
``outside-in'' origin for these events, in which the flare occurs
somewhere outside of the inner accretion disk and eventually
propagates inward to cause a disturbance in the X-rays emitted there
\citep{Eiken2000}.  Simulations of magnetized accretion disks by
\citet{Lovelace} have shown that an increase in the poloidal magnetic
field in the outer part of the disk can lead to an enhanced accretion
rate, which triggers a slightly delayed burst of radiation from the
inner disk.  An alternate idea is that the flare could be produced by
a reverse shock originating within a continuous jet outflow, which
then propagates back down to the inner disk \citep{Eiken2000}.

The possibility that shocks within a preexisting jet could be
responsible for the infrared flares from GRS~1915+105 is intriguing in
light of the variability that we find within each flare.  A model for
internal shocks in microquasar jets was presented by \citet{Kaiser}.
It was primarily used to explain the larger ``class A'' events, but a
modified version might be applicable to the \mbox{$\sim 30$} minute
``class B'' events as well \citep[see][]{Turler}, especially given
that a continuous jet has been observed in GRS~1915+105 and is known
to be associated with the \mbox{$\sim 30$} minute cycles
\citep{Dhawan}.  In addition, \citet{Collins} showed that particularly
large (\mbox{$\sim 300$} mJy) ``class B'' events observed by
\citet{FenderPooley2000} cannot be explained by a simple model in
which the flare results from an increase in the injected particle
density at the base of the jet.

\section{Discussion}

In this section, we outline our scenario for the \mbox{$\sim 30$}
minute variability cycles in GRS~1915+105, based on the observations
presented in this paper.  The key feature of this scenario is that
although GRS~1915+105 always dissipates energy in the jet (via
infrared and radio flares) when the accretion disk becomes active,
this emission is at least partially due to a superposition of smaller
events, one for each oscillation in the X-ray light curve.  Some of
the events in GRS~1915+105 that would previously have been described
as ``class B'' flares (in the sense that there is one large ejection
every \mbox{$\sim 30$} minutes) may really just be a superposition of
smaller ``class C'' flares, which blend together and mimic a larger
event.  Episodes in which a large ejection really {\it does} occur,
leading the infrared flare to dissociate from the X-ray light curve as
the ejected material becomes causally separated from the inner disk
\citep{Eiken98}, seem to require a ``trigger'' spike in the X-ray
light curve at the moment of the state transition, when GRS~1915+105
does not immediately begin the fast X-ray oscillations but rather
enters a phase of duration \mbox{$\sim 300$} seconds in which the
accretion disk is hot and visible but changes on a much slower
timescale.

\subsection{A Continuum of Behavior} \label{continuum}

It is important to realize that the \mbox{$\sim 30$} minute cycles in
GRS~1915+105 do not neatly fall into discrete categories.  From
perusing the figures in this paper, one can easily imagine a continuum
of behavior, from an extreme example in which the initial ejection
completely dominates the infrared flare (Figure \ref{fig:1997}a), to
examples where the initial ejection is weaker and the excess emission
from blended together faint flares is easier to see (Figure
\ref{fig:1997}b), to examples from our 2002 observations where {\it
most} of the infrared emission is due to blended together faint flares
but a small ``peak'' is still observed at the beginning of the event,
perhaps indicating a slightly larger initial ejection (some of the
flares in Figure \ref{fig:lcurve}a), to examples on the other extreme,
where there is no peak at the beginning of the flare and the event is
consistent with the initial ejection being exactly the same as the
subsequent ones (Figure \ref{fig:specfit}b).

In this vein, we point out that X-ray observations which contain a
strong trigger spike \citep[the ``class $\beta$'' state defined
by][]{Belloni2000} have been observed to be accompanied by infrared
flares with a wide range of amplitudes.  The events observed by
\citet{Eiken98} clustered around \mbox{$\sim 100$} mJy but ranged from
\mbox{$\sim 60$} to 200 mJy, while a single event observed by
\citet{Mirabel98} during this X-ray state only reached a peak
amplitude of \mbox{$\sim 30$} mJy.  Meanwhile, the superimposed faint
flares reported in this paper during the ``class $\alpha$'' state in
July 2002 peak relatively consistently at \mbox{$\sim 30$} mJy, but
even these less extreme events may have some variation in amplitude; a
single ``class $\alpha$'' event observed by \citet{Mirabel98} was
accompanied by a flare that only appeared to peak at \mbox{$\sim 10$}
mJy.

The X-ray behavior during the \mbox{$\sim 30$} minute cycles may also
represent a continuum, between cases where there is a clear X-ray
trigger spike and cases where this distinction is not so clear.
Although the initial X-ray oscillation in each 2002 series is very
similar to the subsequent oscillations, it does share a few subtle
characteristics with the 1997 trigger spike.  For example, Figure
\ref{fig:1sec}b shows that the initial 2002 oscillation is
``double-peaked,'' and at the end of the first peak the accretion disk
flux drops sharply while the inner disk temperature continues its
gradual rise.  This is qualitatively the same behavior that was
observed during the 1997 trigger spike (see \S \ref{specevolve}), and
thus the first peak in this oscillation may be regarded as a weak
analog of the trigger spike in 1997, albeit one which leads {\it
immediately} to the X-ray oscillations rather than requiring a
\mbox{$\sim 300$} second phase in which the accretion disk is much
less variable.  Based on these properties, it is not surprising that
the initial 2002 oscillation sometimes appears to be accompanied by a
slightly larger infrared flare (Figure \ref{fig:lcurve}a, as discussed
above) and slightly more variation in the power law index (see \S
\ref{oscillations}) than the oscillations which follow it; it may
really be a weak variant of the 1997 spike which leads to the large
ejections.

\subsection{When Does the Ejection Occur?}

The observations presented in this paper contradict some of the
conclusions of \citet{KleinWolt2002}, who analyzed simultaneous X-ray
and radio observations of GRS~1915+105 and argued that the radio
flares are produced via a continuous ejection of material during the
long, spectrally hard dips in the X-ray light curve \citep[``state C''
in the classification of][]{Belloni2000}.

{\it Independent of any of the specific conclusions we have reached in
this paper, our observations pose a serious challenge for this
argument.}  As is easily visible from Figure \ref{fig:lcurve}, the
duration of each infrared flare in our July 2002 observations is
correlated with the duration of its accompanying period of X-ray
oscillations, not with the previous X-ray dip.  This strongly suggests
that the flares are associated with the X-ray oscillations, while
providing no indication that they are influenced by any properties of
the preceding dip, as might be expected in the model of
\citet{KleinWolt2002}, where a longer dip means that more material is
being ejected into the jet.

The differences which we observe between the 1997 and 2002 infrared
flares also argue against the model of \citet{KleinWolt2002}.  The
X-ray dips in 1997 and 2002 are very similar to each other, and they
are separated by X-ray oscillation cycles which have similar
durations.  The only real difference is that the 2002 dips appear to
be slightly longer (\mbox{$\sim 1000$} seconds as opposed to
\mbox{$\sim 600$} seconds) and to have disk blackbody emission that is
weaker by a factor of \mbox{$\sim 2$} (although it is important to
note that the disk blackbody emission is so weak in both cases that it
is barely detectable).  It is not clear how these slight differences
in the X-ray dips could lead to dramatic differences in the infrared
flares under the \citet{KleinWolt2002} model.  Therefore, the fact
that there appears to have been a large ejection during each
\mbox{$\sim 30$} minute cycle in 1997 but not in 2002 argues for a
scenario in which the flares are associated with properties of the
X-ray state transition (such as the presence or absence of the trigger
spike) and subsequent accretion disk oscillations, which do proceed
much differently in 1997 than in 2002.

We do not dispute the conclusion of \citet{KleinWolt2002} that the
{\it presence} of ``state C'' dips is required for infrared and radio
flares to occur; for example, it is definitely possible that mass
which is later available to be ejected steadily builds up during the
dips \citep[e.g.][]{BelMigFen}.  It is also possible that matter is
being continuously ejected during the dips but that this material does
not directly produced the large flares; instead, the flares could be
produced via internal shocks which occur in the continuous jet when
faster-moving material is ejected at the moment of the state
transition \citetext{\citealp{Vadawale}; see also
\citealp{FenderGallo}}.  The main conclusion of \citet{KleinWolt2002}
which we refute is simply that the flares can be {\it entirely}
explained by the evolution of particles which were ejected during the
dip, without appealing to some other event that occurs afterwards in
the accretion disk.

The above ideas were hinted at by previous multiwavelength
observations containing infrared data
\citep{Eiken98,Mirabel98,Eiken2000}, but they are confirmed by our
present work.  Nonetheless, it will be important to test them with
future radio observations.  The observations analyzed by
\citet{KleinWolt2002} did not contain any examples of the ``class
$\alpha$'' X-ray state for which there was significant overlap in the
X-ray and radio coverage, and in general, there are only a few radio
observations in the literature which overlap with the ``class
$\alpha$'' state.

\subsection{Infrared and Radio Flares} \label{irradioflares}

One of the few previous multiwavelength observations during the
``class $\alpha$'' state raises some questions about our current
results.  \citet{Ueda} obtained \mbox{$\sim 3$} hours of X-ray and
radio (and, separately, X-ray and infrared) observations during this
state.  They detected at least one radio flare that had a {\it
shorter} duration than its accompanying X-ray oscillation period, with
the main part of the flare decaying several hundred seconds before the
end of the oscillations.  This flare was most clearly seen at 1.3 cm
(23 GHz), where it reached an amplitude of \mbox{$\sim 15$} mJy
\citep[Figure 2 of][]{Ueda}.

In most models, radio flares are expected to peak later and last
longer than their associated infrared flares, due to optical depth
effects in the ejected material \citep{vanderLaan,Mirabel98,Collins}.
Therefore, the radio flare observed by \citet{Ueda} does {\it not}
appear to be the counterpart of the superimposed infrared flares we
report in our current work, and, in fact, \citet{Ueda} used their
observations to specifically rule out the possibility that the radio
flares are produced via a superposition of smaller flares associated
with each X-ray oscillation.  In addition, the baseline flux density
of the radio emission during these observations was \mbox{$\sim 10$}
mJy; thus, if there was any ``radio excess'' associated with the X-ray
oscillations after the decay of the flare, it was limited to, at most,
this value.

How can we understand these results?  If the \citet{Ueda} observation
is representative of the radio behavior during the ``class $\alpha$''
state, then the initial ejections at the moment of the state
transition may really be distinct, even though they do not appear that
way in the infrared light curve.  The initial ejection might have a
relatively flat infrared-to-radio spectrum (and thus a prominent radio
signature), while the repeating events associated with each X-ray
oscillation could have little associated radio emission.  If this is
the case, then the differences between the ``class $\alpha$'' and
``class $\beta$'' states are not really as strong as we have indicated
in this paper; both might produce ejections at the moment of the state
transition which are substantially different from the subsequent ones
(although it still the case that the ``class $\beta$'' events produce,
on average, much larger infrared and radio flares than the ``class
$\alpha$'' events).

Alternatively, it is possible that the \citet{Ueda} event was not
representative of the normal behavior during the ``class $\alpha$''
state.  We have analyzed the X-ray light curve and spectrum from this
event and found that there is a relatively sharp double peak in the
first oscillation in this series, whose first peak has spectral and
temporal characteristics particularly reminiscent of a ``class
$\beta$'' trigger spike (see \S \ref{continuum}).  In addition, there
is an unusual, \mbox{$\sim 10$} second spike in the X-ray light curve
which occurs \mbox{$\sim 30$} seconds {\it before} the state
transition (although it is spectrally different from the trigger spike
and appears to simply be a thermal ``bump'' in the spectrum).  This
behavior suggests that \citet{Ueda} may have observed an anomalous
event, in which the initial ejection was much different than those
which normally occur during the ``class $\alpha$'' state.

Furthermore, the only other ``class $\alpha$'' radio observation we
could find in the literature appears to be consistent with the
interpretation we have put forth in this paper.  Figure 1 of
\citet{Mirabel98} shows that during the ``class $\alpha$'' state, the
flare duration increases with wavelength as expected, including one
case in which the flare was simultaneously observed in the infrared
and three different radio wavelengths.  The X-ray oscillation period
associated with this event was not observed, but it is constrained to
have been within a short (\mbox{$\sim 30$} minute) gap in the X-ray
coverage that overlaps with the infrared flare and is of significantly
shorter duration than the radio flares.  Thus, it is broadly
consistent with our current observations.

The ultimate answer to the questions raised here may rest on how much
variation there is in the amplitude of the ``class C'' flares
associated with each X-ray oscillation.  If the ``radio excess'' seen
in the \citet{Ueda} observations (\mbox{$< 10$} mJy if it exists at
all) is typical, then it is difficult to see how the repeating, larger
radio flares during the ``class $\alpha$'' state \citep[see \S
\ref{origin}, as well as][]{Mirabel98}, could be explained without
resorting to initial ejections that are different from the subsequent
events.  Also note that the ``class $\rho$'' state in GRS~1915+105,
which consists of an extremely long series of oscillations very
similar to those seen during the ``class $\alpha$'' state
\citep{Belloni2000}, is associated with steady radio emission of only
\mbox{$\sim 4$} mJy at 15 GHz \citep{KleinWolt2002}.

To further complicate the picture, the infrared observations of
\citet{Ueda} show at least one instance (and likely more) in which a
{\it delayed} \mbox{$\sim 30$} minute infrared flare is seen during
the ``class $\alpha$'' state, beginning at the end of its associated
X-ray oscillation period rather than the beginning.  (Note that as in
our observations, the duration of this flare still matches the
duration of the oscillations.)  It is unclear what could have caused
this unusual delay, which is inconsistent with all the infrared flares
we report in this paper.  \citet{Ueda} suggested the possibility of an
internal shock occurring far out along the jet, but it is a puzzling
event under any interpretation.

Clearly, the only way to resolve the issues raised here is to obtain
simultaneous X-ray, infrared and radio observations over a long
stretch of events.  In the absence of such observations, we can only
point out that the conclusions we have reached in this paper apply to
all \mbox{$\sim 15$} simultaneous episodes that we observed.  Overall,
it is clear that the 1997 ``class $\beta$'' events produce
qualitatively different flaring behavior than the 2002 ``class
$\alpha$'' events and that these differences are related to the
different ways in which the state transition proceeds and in
particular to the X-ray trigger spike, which appears most strongly in
the ``class $\beta$'' observations.  However, the exact manner in
which properties of the trigger spike control the behavior of the
initial flare --- for example, whether the ``sharpness'' of the spike
or the delay between it and the subsequent oscillations is the most
important feature --- cannot yet be conclusively determined.

\section{Conclusions}

We have presented simultaneous infrared and X-ray observations of the
microquasar GRS~1915+105 (at 1-second time resolution) during a period
of jet ejection in July 2002.  By comparing these observations to
those obtained in August 1997 by \citet{Eiken98}, we arrive at a
picture in which a large ejection may not always be present during the
\mbox{$\sim 30$} minute state transitions in GRS~1915+105.  The
observed infrared flares may sometimes entirely be a superposition of
smaller, more complex phenomena, with large, infrared-bright ejections
superimposed on them only when there is a ``trigger'' spike in the
X-ray light curve.  In particular, we find that:
\begin{enumerate}

\item The duration of each infrared flare in 2002 matches the duration
of its accompanying X-ray oscillation period.

\item There is one instance in which a single, isolated X-ray
oscillation occurs in the light curve, accompanied by a faint infrared
``subflare'' superimposed on one of the main flares.

\item These observations are consistent with a scenario in which each
X-ray oscillation has an associated faint infrared flare and these
flares blend together to form, and entirely comprise, the larger
events.

\item The main difference between our X-ray observations and the
August 1997 observations is the presence of a sharp ``trigger'' spike
in the 1997 X-ray light curve during each \mbox{$\sim 30$} minute
cycle, which occurs \mbox{$\sim 300$} seconds before the accretion
disk oscillations and signals a sharp change in the spectral
properties of the source.

\item The trigger spike appears to be associated with initial, large
infrared flares, which are observed in 1997 but not in 2002.

\item The 1997 X-ray oscillations contain dramatic changes in the high
energy power law component of the X-ray spectrum, which are not
observed during the 2002 oscillations.  This is somewhat puzzling,
since both sets of oscillations produce similar infrared subflares.

\item There is tentative evidence that the infrared subflare observed
in 2002 leads its associated X-ray oscillation by \mbox{$\sim 30$
seconds}, possibly suggesting an ``outside-in'' origin for these
events, as in observations by \citet{Eiken2000}.

\item It is unlikely that any of the 2002 flares can be explained
solely by the continuous ejection of material during the spectrally
hard X-ray ``dip'' which precedes each set of X-ray oscillations, as
suggested by \citet{KleinWolt2002}.

\end{enumerate}

Although many authors have speculated on the importance of the trigger
spike as a signature of large jet ejections
\citep[e.g.][]{Eiken98,Mirabel98,yadav,FenderGallo}, the observations
presented in this paper confirm it.  Our 2002 X-ray observations,
which have many similarities to the \citet{Eiken98} observations
except for the absence of a distinctive trigger spike, do not show the
same large flares in the infrared.  Thus, these data are important in
allowing us to pinpoint which aspects of the X-ray behavior are
related to which types of events in the jet.

Overall, we find that jet production in GRS~1915+105 is a complex,
unsteady phenomenon, down to the fastest timescales that have been
observed.  The jet and the accretion disk appear to interact on a
regular basis, perhaps every time an oscillation occurs within the
inner portion of the disk.  These details need to be taken into
account in accretion disk models, such as in the simulations of
radiation pressure dominated disk evolution by \citet{Janiuk}, which
found that energy dissipation in a corona or via an outflow from the
disk may be important in reproducing the GRS~1915+105 X-ray emission.

\acknowledgments

We thank the staff at Palomar Observatory and the members of the {\it
Rossi X-Ray Timing Explorer} team for their help with these
observations.  We also thank G. Pooley for providing the radio data
from the Ryle Telescope, and R. Lovelace, M. Tagger and the anonymous
referee for useful discussions and suggestions.  This research has
made use of NASA's Astrophysics Data System, as well as the High
Energy Astrophysics Science Archive Research Center (HEASARC) provided
by NASA's Goddard Space Flight Center.  D.~M.~R. is supported by a
National Science Foundation Graduate Research Fellowship.  S.~S.~E. is
supported in part by an NSF CAREER award (NSF-9983830).


\begin{thebibliography}{}

\bibitem[Abramowicz et al.(1988)]{Abramowicz} Abramowicz, M. A.,
Czerny, B., Lasota, J. P., \& Szuszkiewicz, E. 1988, \apj, 332, 646

\bibitem[Belloni et al.(1997a)]{Belloni97a} Belloni, T., M\'{e}ndez,
M., King, A. R., van der Klis, M., \& van Paradijs, J. 1997a, \apj,
479, L145

\bibitem[Belloni et al.(1997b)]{Belloni97b} Belloni, T., M\'{e}ndez,
M., King, A. R., van der Klis, M., \& van Paradijs, J. 1997b, \apj,
488, L109

\bibitem[Belloni et al.(2000)]{Belloni2000} Belloni, T., Klein-Wolt,
M., M\'{e}ndez, M., van der Klis, M., \& van Paradijs, J. 2000, \aap,
355, 271

\bibitem[Belloni, Migliari \& Fender(2000)]{BelMigFen} Belloni, T.,
Migliari, S., \& Fender, R. P. 2000, \aap, 358, L29

\bibitem[Castro-Tirado et al.(1994)]{Castro} Castro-Tirado, A. J.,
Brandt, S., Lund, N., Lapshov, I., Sunyaev, R. A., Shlyapnikov, A. A.,
Guziy, S., \& Pavlenko, E. P. 1994, \apjs, 92, 469

\bibitem[Chapuis \& Corbel(2004)]{Chapuis} Chapuis, C., \& Corbel,
S. 2004, \aap, 414, 659

\bibitem[Collins, Kaiser \& Cox(2003)]{Collins} Collins, R. S.,
Kaiser, C. R., \& Cox, S. J. 2003, \mnras, 338, 331

\bibitem[Dhawan, Mirabel \& Rodr\'\i guez(2000)]{Dhawan} Dhawan, V.,
Mirabel, I. F., \& Rodr\'\i guez, L. F. 2000, \apj, 543, 373

\bibitem[Done, Wardzi\'{n}ski \& Gierli\'{n}ski(2004)]{Done2004} Done,
C., Wardzi\'{n}ski, G., \& Gierli\'{n}ski, M. 2004, \mnras, 349, 393

\bibitem[Eikenberry et al.(1998a)]{Eiken98} Eikenberry, S. S.,
Matthews, K., Morgan, E. H., Remillard, R. A., \& Nelson, R. W. 1998a,
\apj, 494, L61

\bibitem[Eikenberry et al.(1998b)]{Eiken98b} Eikenberry, S. S.,
Matthews, K., Murphy, T. W., Jr., Nelson, R. W., Morgan, E. H.,
Remillard, R. A., \& Muno, M. 1998b, \apj, 506, L31

\bibitem[Eikenberry et al.(2000)]{Eiken2000} Eikenberry, S. S.,
Matthews, K., Muno, M., Blanco, P. R., Morgan, E. H., \& Remillard,
R. A. 2000, \apj, 532, L33

\bibitem[Eikenberry \& van Putten(2003)]{VanPutten} Eikenberry, S. S.,
\& van Putten, M. H. H. M. 2003, \apj, submitted (astro-ph/0304386)

\bibitem[Fender et al.(1997)]{Fender97} Fender, R. P., Pooley, G. G.,
Brocksopp, C., \& Newell, S. J. 1997, \mnras, 290, L65

\bibitem[Fender \& Pooley(1998)]{FenderPooley98} Fender, R. P., \&
Pooley, G. G. 1998, \mnras, 300, 573

\bibitem[Fender et al.(1999a)]{Fender99} Fender, R. P., Garrington,
S. T., McKay, D. J., Muxlow, T. W. B., Pooley, G. G., Spencer, R. E.,
Stirling, A. M., \& Waltman, E. B. 1999a, \mnras, 304, 865

\bibitem[Fender et al.(1999b)]{FenderCompton} Fender, R., et
al. 1999b, \apj, 519, L165

\bibitem[Fender \& Pooley(2000)]{FenderPooley2000} Fender, R. P., \&
Pooley, G. G. 2000, \mnras, 318, L1

\bibitem[Fender(2004)]{Fender2004} Fender, R. P. 2004, in Compact
Stellar X-Ray Sources, ed. W. H. G. Lewin, \& M. van der Klis
(Cambridge: Cambridge University Press), in press (astro-ph/0303339)

\bibitem[Fender \& Belloni(2004)]{FenderBelloni} Fender, R., \&
Belloni, T. 2004, \araa, 42, 317

\bibitem[Fender, Belloni \& Gallo(2004)]{FenderGallo} Fender, R. P.,
Belloni, T. M., \& Gallo, E. 2004, \mnras, 355, 1105

\bibitem[Feroci et al.(1999)]{Feroci} Feroci, M., Matt, G., Pooley,
G., Costa, E., Tavani, M., \& Belloni, T. 1999, \aap, 351, 985

\bibitem[Fuchs, Mirabel \& Claret(2003)]{fuchs} Fuchs, Y., Mirabel,
I. F., \& Claret, A. 2003, \aap, 404, 1011

\bibitem[Greiner, Morgan \& Remillard(1996)]{Greiner96} Greiner, J.,
Morgan, E. H., \& Remillard, R. A. 1996, \apj, 473, L107

\bibitem[Greiner, Cuby \& McCaughrean(2001)]{GreinerMass} Greiner, J.,
Cuby, J. G., \& McCaughrean, M. J. 2001, Nature, 414, 522

\bibitem[Greiner et al.(2001)]{GreinerComp} Greiner, J., Cuby, J. G.,
McCaughrean, M. J., Castro-Tirado, A. J., \& Mennickent, R. E. 2001,
\aap, 373, L37

\bibitem[Janiuk, Czerny \& Siemiginowska(2002)]{Janiuk} Janiuk, A.,
Czerny, B., \& Siemiginowska, A. 2002, \apj, 576, 908

\bibitem[Kaiser, Sunyaev \& Spruit(2000)]{Kaiser} Kaiser, C. R.,
Sunyaev, R., \& Spruit, H. C. 2000, \aap, 356, 975

\bibitem[Klein-Wolt et al.(2002)]{KleinWolt2002} Klein-Wolt, M.,
Fender, R. P., Pooley, G. G., Belloni, T., Migliari, S., Morgan,
E. H., \& van der Klis, M. 2002, \mnras, 331, 745

\bibitem[Livio, Pringle \& King(2003)]{Livio} Livio, M., Pringle,
J. E., \& King, A. R. 2003, \apj, 593, 184

\bibitem[Lovelace, Romanova \& Newman(1994)]{Lovelace} Lovelace,
R. V. E., Romanova, M. M., \& Newman, W. I. 1994, \apj, 437, 136

\bibitem[Markoff, Falcke \& Fender(2001)]{Markoff} Markoff, S.,
Falcke, H., \& Fender, R. 2001, \aap, 372, L25

\bibitem[Markoff et al.(2003)]{Markoff2003} Markoff, S., Nowak, M.,
Corbel, S., Fender, R., \& Falcke, H. 2003, \aap, 397, 645

\bibitem[McClintock \& Remillard(2004)]{McClintock} McClintock, J. E.,
\& Remillard, R. A. 2004, in Compact Stellar X-Ray Sources,
ed. W. H. G. Lewin, \& M. van der Klis (Cambridge: Cambridge
University Press), in press (astro-ph/0306213)

\bibitem[Merloni, Fabian \& Ross(2000)]{Merloni2000} Merloni, A.,
Fabian, A. C., \& Ross, R. R. 2000, \mnras, 313, 193

\bibitem[Migliari \& Belloni(2003)]{migliari} Migliari, S., \&
Belloni, T. 2003, \aap, 404, 283

\bibitem[Mirabel \& Rodr\'\i guez(1994)]{Mirabel94} Mirabel, I. F., \&
Rodr\'\i guez, L. F. 1994, Nature, 371, 46

\bibitem[Mirabel et al.(1994)]{Mirabeletal94} Mirabel, I. F., et
al. 1994, \aap, 282, L17

\bibitem[Mirabel et al.(1998)]{Mirabel98} Mirabel, I. F., Dhawan, V.,
Chaty, S., Rodr\'\i guez, L. F., Mart\'\i, J., Robinson, C. R., Swank,
J., \& Geballe, T. R. 1998, \aap, 330, L9

\bibitem[Mirabel \& Rodr\'\i guez(1999)]{Mirabel99} Mirabel, I. F., \&
Rodr\'\i guez, L. F. 1999, \araa, 37, 409

\bibitem[Mitsuda et al.(1984)]{Mitsuda} Mitsuda, K., et al. 1984,
\pasj, 36, 741

\bibitem[Muno, Morgan \& Remillard(1999)]{Muno1999} Muno, M. P.,
Morgan, E. H. \& Remillard, R. A. 1999, \apj, 527, 321

\bibitem[Muno et al.(2001)]{Muno2001} Muno, M. P., Remillard, R. A.,
Morgan, E. H., Waltman, E. B., Dhawan, V., Hjellming, R. M., \&
Pooley, G. 2001, \apj, 556, 515

\bibitem[Persson et al.(1998)]{Persson} Persson, S. E., Murphy, D. C.,
Krzeminski, W., Roth, M., \& Rieke, M. J. 1998, \aj, 116, 2475

\bibitem[Pooley \& Fender(1997)]{PooleyFender97} Pooley, G. G., \&
Fender, R. P. 1997, \mnras, 292, 925

\bibitem[Poutanen(1998)]{Poutanen} Poutanen, J. 1998, in Theory of
Black Hole Accretion Disks, ed. M. A. Abramowicz, G. Bj\"{o}rnsson, \&
J. E. Pringle (Cambridge: Cambridge University Press), 100

\bibitem[Remillard et al.(2003)]{RemillardCargese} Remillard, R.,
Muno, M., McClintock, J., \& Orosz, J. 2003, in Proceedings of the
Fourth Microquasar Workshop, New Views on Microquasars,
ed. P. Durouchoux, Y. Fuchs, \& J. Rodriguez (Kolkata: Center for
Space Physics), 57 (astro-ph/0208402)

\bibitem[Rodriguez et al.(2002)]{Rodriguez} Rodriguez, J.,
Varni\`{e}re, P., Tagger, M., \& Durouchoux, P. 2002, \aap, 387, 487

\bibitem[Rodr\'\i guez \& Mirabel(1999)]{RodMirabel99} Rodr\'\i guez,
L. F., \& Mirabel, I. F. 1999, \apj, 511, 398

\bibitem[Rothstein \& Eikenberry(2003)]{Rothstein} Rothstein, D. M.,
\& Eikenberry, S. S. 2003, in Proceedings of the Fourth Microquasar
Workshop, New Views on Microquasars, ed. P. Durouchoux, Y. Fuchs, \&
J. Rodriguez (Kolkata: Center for Space Physics), 341
(astro-ph/0208068)

\bibitem[Taam, Chen \& Swank(1997)]{taam} Taam, R. E., Chen, X., \&
Swank, J. H. 1997, \apj, 485, L83

\bibitem[Tagger et al.(2004)]{Tagger} Tagger, M., Varni\`{e}re, P.,
Rodriguez, J., \& Pellat, R. 2004, \apj, 607, 410

\bibitem[T\"{u}rler et al.(2004)]{Turler} T\"{u}rler, M., Courvoisier,
T. J.-L., Chaty, S., \& Fuchs, Y. 2004, \aap, 415, L35

\bibitem[Ueda et al.(2002)]{Ueda} Ueda, Y., et al. 2002, \apj, 571, 918

\bibitem[Vadawale et al.(2003)]{Vadawale} Vadawale, S. V., Rao, A. R.,
Naik, S., Yadav, J. S., Ishwara-Chandra, C. H., Pramesh Rao, A., \&
Pooley, G. G. 2003, \apj, 597, 1023

\bibitem[van der Laan(1966)]{vanderLaan} van der Laan, H. 1966,
Nature, 211, 1131

\bibitem[Watarai et al.(2000)]{Watarai2000} Watarai, K., Fukue, J.,
Takeuchi, M., \& Mineshige, S. 2000, \pasj, 52, 133

\bibitem[Watarai \& Mineshige(2003)]{Watarai2003} Watarai, K., \&
Mineshige, S. 2003, \apj, 596, 421

\bibitem[Yadav(2001)]{yadav} Yadav, J. S. 2001, \apj, 548, 876

\end{thebibliography}
\end{document}